Pavlov, Oleg V.; Smirnova, Natalia V.; Smirnova, Elena V.

**Working Paper**

# Enhancing Economic Literacy through Causal Diagrams

# Enhancing Economic Literacy through Causal Diagrams

Oleg V. Pavlov[1*], Natalia V. Smirnova[2], and Elena V. Smirnova[3]

## Abstract

A literacy-targeted approach to economic instruction draws on insights from cognitive science. It highlights that students process complex economic information by constructing and modifying schemas that represent economic material. Following this approach, we developed a set of instructional activities centered around *causal diagrams* that promote a deeper understanding of economic topics beyond the traditional lecture-based methods. Our results show that *structural debriefing* activities can be used effectively to introduce students to the causal diagrams that explain key economic relationships in the national income model, government-purchases multiplier and tax multiplier.

## Introduction

Goffe and Wolla (2024) wrote in support of a *literacy-targeted approach* to economic instruction, which emphasizes in-depth coverage of fewer economic topics compared to a typical economics course. They draw on several concepts from cognitive and learning sciences, including schemas, constructivism and deliberate practice. Research indicates that individuals arrange information into schemas, which are "networks of ideas, concepts and procedures" (Goffe and Wolla, 2024: p. 157). Student comprehension and retention of complex models and topics depend on such schemas representing economic material. Constructivism, the second concept, posits that students build knowledge into a complex system, continuously adding new ideas to a set of preexisting constructs. In other words, students process complex economic information by constructing and modifying schemas (Ambrose et al., 2010). As novices become experts, they develop and improve their schemas. The concept of deliberate practice refers to instructional activities that are "specifically designed to maximize schema formation" (Goffe and Wolla 2024: 159).

[1] Social Science and Policy Studies, Worcester Polytechnic Institute, USA & GLO, opavlov@wpi.edu
[2] Department of Economics, University of Connecticut, Stamford, CT, USA, natalia.smirnova@uconn.edu
[3] Department of Management, Marketing and Finance, School of Business, State University of New York at Old Westbury, Old Westbury, NY, USA, smirnovae@oldwestbury.edu
[*] Corresponding author

The authors are grateful for the helpful feedback from Mary Suiter and Scott Wolla, and the comments of the participants of the 2024 Professors' Conference at the St. Louis Federal Reserve.

Following the literacy-targeted approach to economic instruction, we designed and implemented a set of pedagogical activities that help students with constructing their economic schemas. These activities involve students working with *causal diagrams* through a process called *structural debriefing*. Structural debriefing has been used to improve clarity and learning effectiveness of instructional simulations (Capelo and Silva 2020; Capelo et al. 2021; Capelo et al. 2024; Pavlov et al. 2015; Qudrat-Ullah 2020) and educational videogames (Kim and Pavlov 2019). We adapt structural debriefing for economic instruction by using it to explain the causal relationships discussed in macroeconomic textbooks. This involves visualizing economic variables and their interrelationships as graphical diagrams. Each activity can be integrated into traditional lecture-based instruction. These causal diagrams supplement the textual information and traditional graphs found in modern economic textbooks, providing an additional type of graph beyond the ubiquitous supply-and-demand graphs.

We employ the *design-based approach* (Collins et al. 2004; McKenney and Reeves 2012) to develop and iteratively improve these debriefing activities in response to in-class observations, student performance, and feedback from students. The goal has been to develop lesson plans and course materials that can be used by economics instructors who are not experts in causal diagrams.

In the following sections, we explain causal diagrams, our research method, and the design process. Then we describe how we implemented the structural debriefing activities and the findings from the classroom. The final sections offer a discussion and conclusion.

## Causal diagrams

Research shows that students frequently underestimate the complexity of causality, thinking in simple linear terms, like "A causes B" (Perkins & Grotzer, 2005). However, real-world systems are much more complex, consisting of numerous interacting parts that create intricate causal networks with mutual causality, feedback loops, domino effects, and spiraling causality (Grotzer, 2012). Several cognitive factors limit our causal understanding of the world. For example, individuals who observe only direct responses to actions fail to recognize the entire causal complexity of the system. Additionally, people often mistake statistical correlations for true causation (Sterman 2010). And in the quest for efficiency, our minds overlook a lot of useful information, as certain details simply do not capture our attention. Moreover, introductory economics courses often cover so much material that it overwhelms students (Goffe and Wolla 2024).

A causal diagram is an effective yet simple tool for representing causal information (Pearl and Mackenzie 2018: 39). Common in systems thinking, system dynamics (Maani and Cavana, 2007; Sterman, 2000) and artificial intelligence (Pearl and Mackenzie 2018), these diagrams visually represent the cause-and-effect relationships between variables. Variables form nodes, and arrows indicate causal relationships. Positive arrows show that cause and effect move in the same direction, while negative arrows indicate that the cause and effect are inversely related. An example in Figure 1 demonstrates how to use this notation to visualize the two effects of price on revenue:

- Revenue generated by each item sold increases with price, which is shown as a positive arrow. The greater marginal revenue contributes to overall revenue, hence a positive arrow between these two variables. This causal chain of two positive arrows captures the combined positive effect of higher prices on revenue due to the increased revenue per item sold.

- Additionally, a higher price depresses sales, hence a negative arrow between these two variables. Because sales and revenue move in the same direction, the arrow between sales and revenue is positive. A negative arrow followed by a positive arrow form a negative causal chain that captures the negative effect of higher prices on revenue due to a decline in sales.

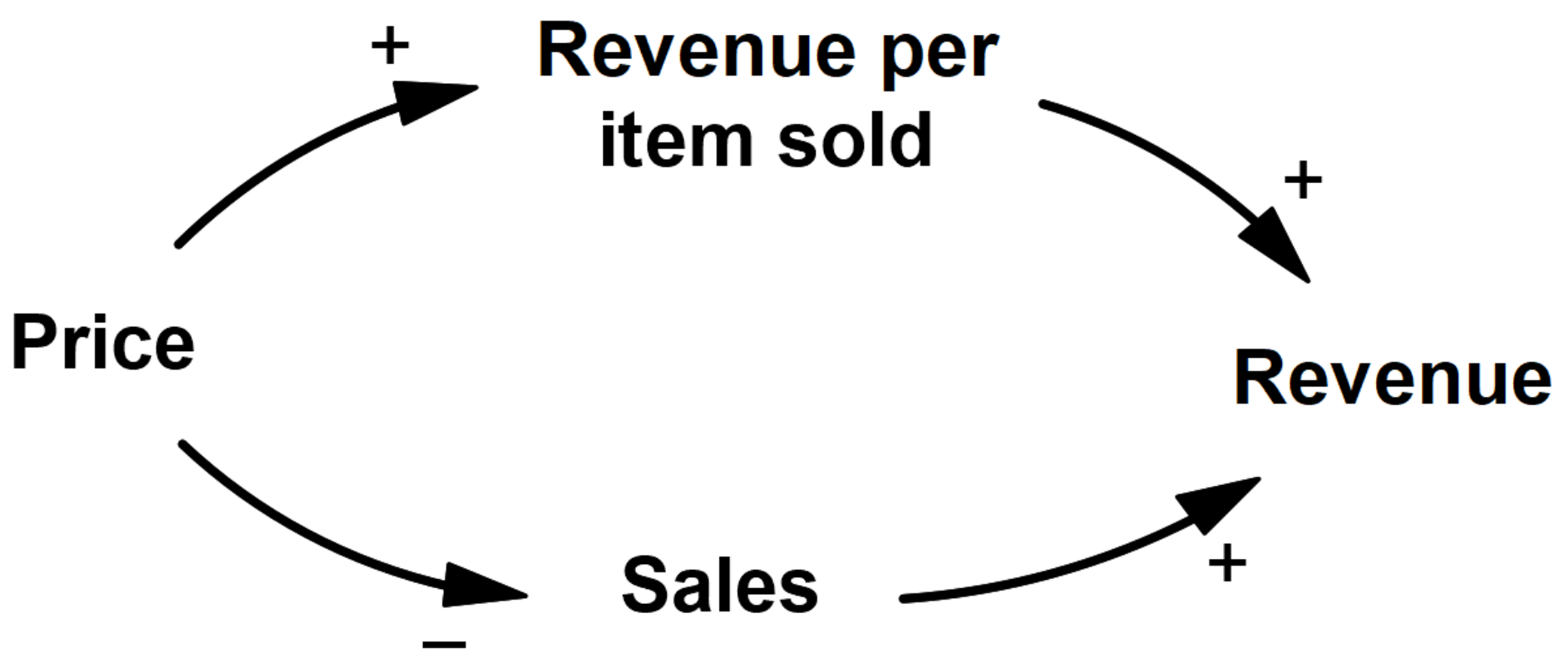

**Figure 1:** An example of a causal diagram for the effect of price on revenue. The diagram shows the positive effect (the upper causal chain) and the negative effect (through sales).

The effectiveness of causal diagrams in an economics classroom was studied in a series of articles by Jägerskog and colleagues (Jägerskog et al., 2019; Jägerskog, 2021b; Jägerskog, 2021a) who found that with causal diagrams students achieve a deeper understanding of the economic material than with the traditional supply and demand graph alone.

## Research method

This research studies the use of causal diagrams in the classroom with the purpose of helping students develop economic schemas for deeper understanding of the class material. We develop and implement three structural debriefing activities that guide students through the process of constructing economic schemas by working with causal diagrams. This section explains structural debriefing and the design of this study.

## Structural debriefing

A debriefing is an activity for reviewing and analyzing an event, like a business simulation, a military operation, or a surgery, with the participants who experienced the event (Crookall 2010; Lederman 1992). A debriefing may include a discussion and journal writing, some type of analysis, personalization, and generalization of the experience (Lederman 1992; Petranek et al. 1992; Steinwachs 1992).

Structural debriefing is a type of debriefing during which participants identify key variables that are relevant to an event, discuss the causal relationships between the variables, and connect system behavior to its causal structure (Pavlov et al. 2015). Structural debriefing helps to improve performance and learning in instructional management simulations (Capelo and Silva 2020; Capelo et al. 2021; Capelo et al. 2024; Qudrat-Ullah 2020) and educational video games (Kim and Pavlov 2019; Pavlov et al. 2019). The goal of structural debriefing is to understand the causal structure of a particular economic situation, effect or mechanism. For example, Capelo et al. (2021) found that students who participated in debriefing sessions about the cause-and-effect relationships in a business venture simulation and discussed the relationship between the structure of the simulated system and its behavior, performed better and had a better understanding of the model dynamics. Similarly, Qudrat-Ullah (2020) demonstrated that in a fishery management simulation, understanding the causal structure of a complex task allows students to develop better heuristics and improve their decision-making.

In this study, structural debriefing helps students recognize key variables and causal connections within the models and concepts covered in a typical macroeconomics textbook. As a literacy-targeted activity, structural debriefing guides students through the explicit construction of causal diagrams, which serve as schemas encompassing the economic material. A notable feature of causal diagrams is their adaptability: as students cover additional material in class, new variables and causal relationships can be added to the existing diagrams, thereby integrating new knowledge with pre-existing constructs.

## Study design

The goal of our investigation is to develop a pedagogical framework that supports literacy-targeted economic instruction by leveraging the affordances of *causal diagrams* as tools for building coherent and inter-connected schemas. We integrate causal diagrams into the curriculum through *structural debriefing* activities. Given the novelty of this approach, we employ a design-based methodology (Collins et al. 2004; McKenney and Reeves 2012) to conduct formative research aimed at validating and refining the structural debriefing activities for literacy-targeted instruction in economics education. The design-based approach allows us to adapt general principles to specific situations and reflect on our findings. By following this approach, we can identify potential challenges, gaps, and opportunities during the development phase. After implementing each curriculum activity, we discuss the classroom observations and analyze student performance and feedback. Based on these data, we rapidly iterate on the materials for subsequent activities.

The research question of this study is:

> *RQ: How can instructors utilize causal diagrams and structural debriefing activities for literacy-targeted instruction in a macroeconomics classroom?*

This study involved the following steps:

> *Step 1*: Develop causal diagrams for three topics from a mainstream macroeconomics textbook.
>
> *Step 2*: Prepare structural debriefing materials for each topic that an instructor can use in the classroom.
>
> *Step 3*: Revise the activities based on the classroom experience, student performance, and student feedback.

Over several months, we discussed the design principles and prepared the first drafts of causal diagrams and teaching materials, which then were continuously improved as the study progressed.

## Design process and products

In our initial discussions, we considered the design principles for lesson plans that could be used by economics instructors with no prior experience with causal diagrams. As we explored the affordances and limitations of the structural debriefing protocol, we identified constraints such as limited classroom time and the instructors' unfamiliarity with causal diagrams. We agreed that our approach needed to incorporate the following design principles:

> ***Develop self-contained activities for macroeconomic topics:*** For this study, we relied on the 9th edition of Mankiw's "Macroeconomics," a popular intermediate-level textbook. While the textbook's material transitions from one chapter to the next, allowing new variables and causal relationships to be added to the existing diagrams, we decided to limit each causal diagram to one topic to keep things simple. We also decided to develop activities as self-contained curricular units to give instructors the flexibility to choose which activities to include in their courses.
>
> ***Simplify tasks for students:*** Although structural debriefing can start with a blank page, we recognized that asking students to identify key variables and arrange them into causal diagrams might be overwhelming, especially for those unfamiliar with causal diagrams and with limited time. To simplify the tasks, we decided that each activity would include '*causal skeletons*', which are graphs with variables and undirected edges between them.
>
> ***Provide instructors with lesson plans, slides and assignments:*** To minimize barriers to adopting this curriculum, we decided to develop detailed lesson plans, slides, and assignments for each activity.

After reviewing the syllabus for an Intermediate Macroeconomics course taught by one of the authors, we decided to prepare debriefing activities for the following three topics: the national income model, the government-purchases multiplier, and the tax multiplier. As Step 1 of the study, we developed causal diagrams for each activity, as presented below.

**Activity 1 – The national income model**: We developed a causal diagram (Figure 2) for the key variables of the national income model, which is covered in two chapters in Mankiw (2016). The diagram consists of 24 variables and 32 causal links. Without this visual aid, students would need to keep these elements in their "working memory" (Goffe and Wolla 2024) – not an easy task for any student. To read this diagram, we can start, for example, by focusing on the variable national output, Y, which depends on three other variables: available technology, capital, K, and employed labor, L. As available technology, capital, or labor increases, so does national output. These positive causal relationships are shown as positive arrows leading to national output. National output determines national income, indicated by a positive arrow from output to income. The proportional relationship between disposable income and national income is shown as a positive arrow. However, if taxes increase, disposable income drops, represented by a negative arrow from taxes to disposable income. Since national savings is the sum of public and private savings, we show this with two positive arrows leading from public savings and private savings to national savings. The rest of the diagram can be understood similarly by moving from one variable to the next. The causal diagram helps students keep track of the variables and relationships in the national income model. We also experimented with using color to group related variables and relationships. For example, elements related to the financial system are blue.

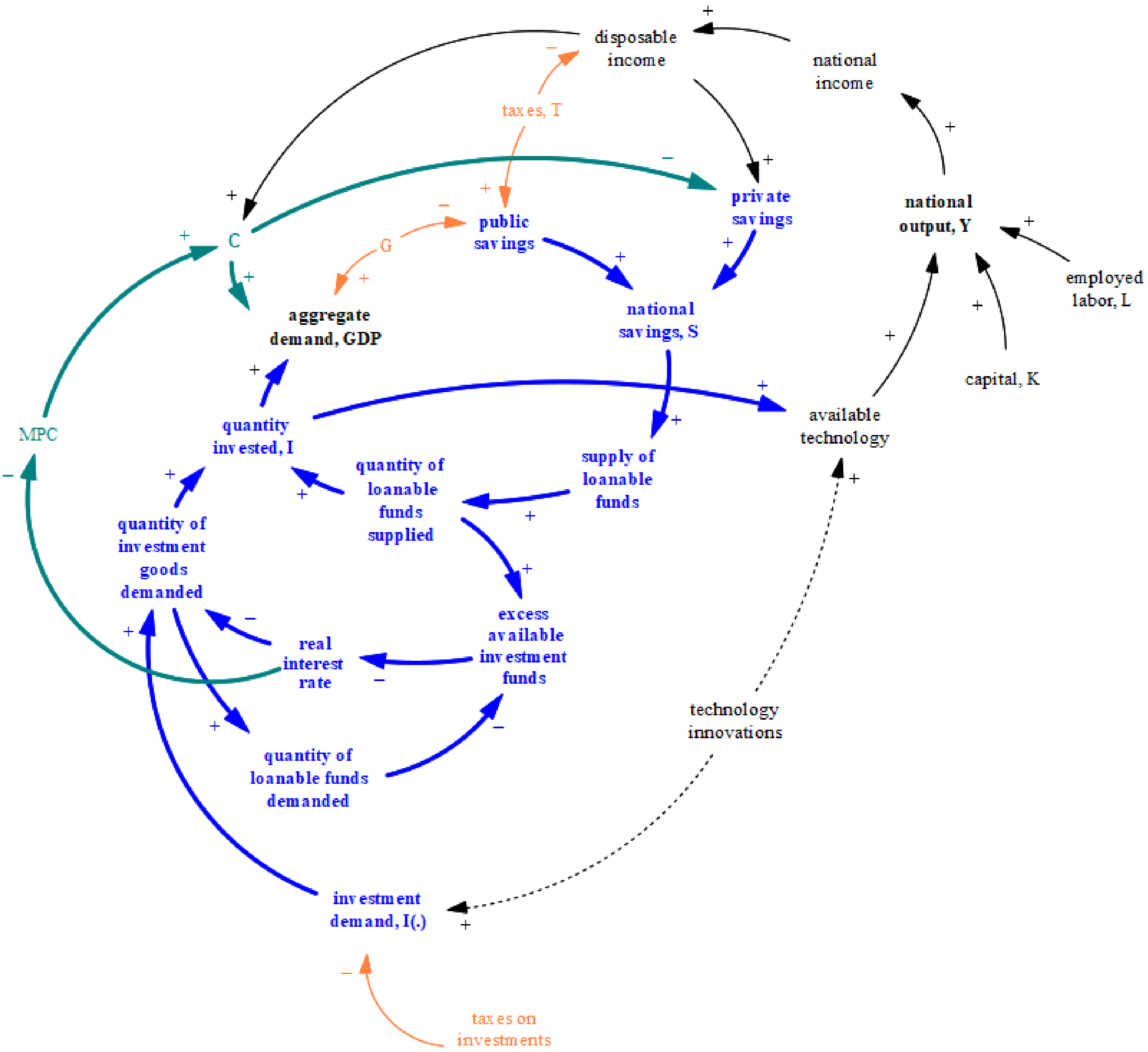

**Figure 2:** The causal diagram of the national income model.

**Activity 2 – Government-purchases multiplier:** A causal diagram in Figure 3 was prepared to explain the government-purchases multiplier in Mankiw's chapter on Aggregate Demand. The diagram includes eight key variables and eight causal connections between them. There is only one negative arrow, which represents the relationship between income tax, T, and disposable income, Y – T. The circular causal connections form a positive feedback loop, indicated by a circular arrow with a plus sign in the middle. This feedback is explicitly mentioned in Mankiw as the mechanism that leads to the government-purchases multiplier effect. The causal diagram helps students visualize this feedback. Due to the relative simplicity of this diagram, we did not color-code the variables.

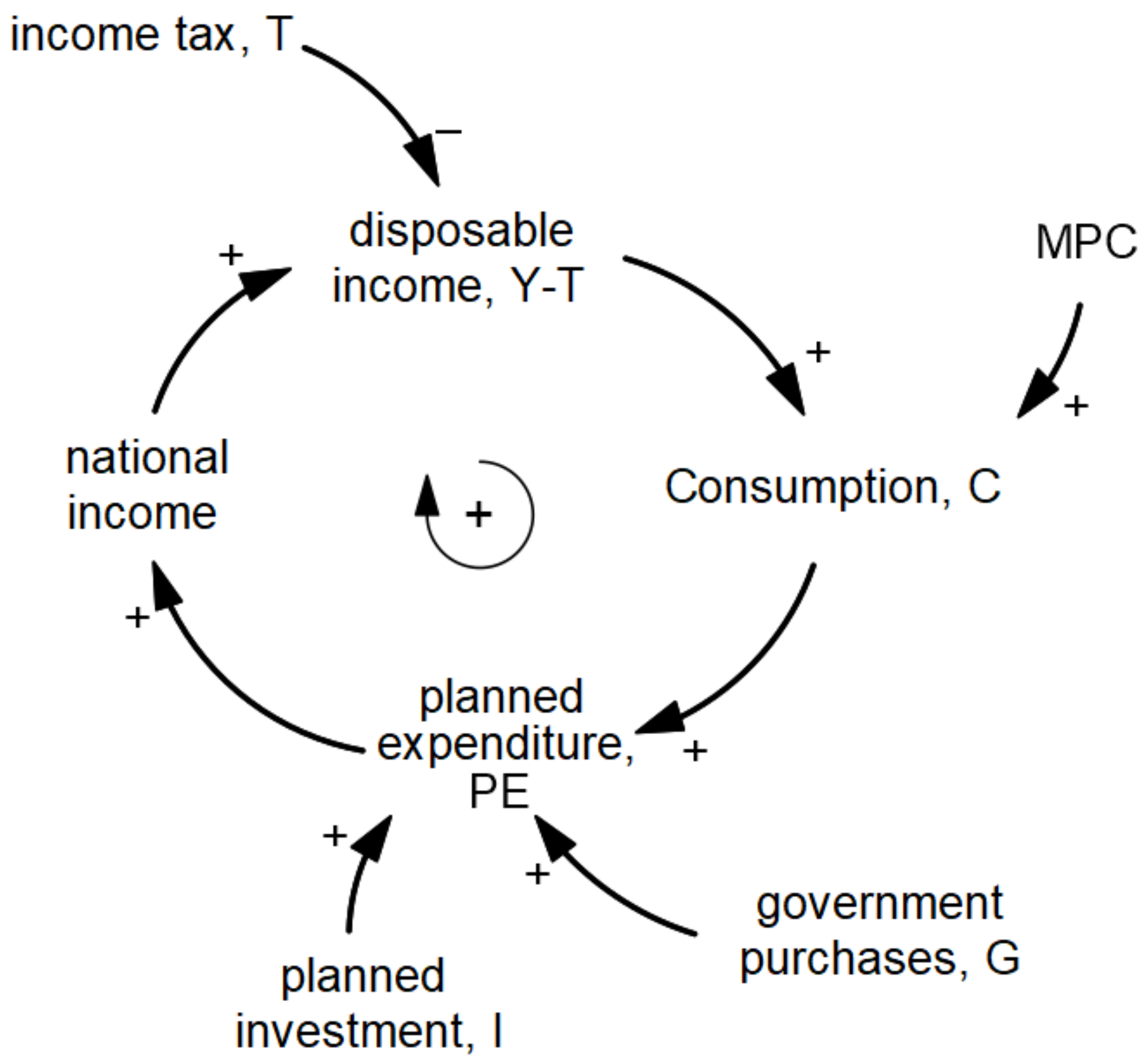

**Figure 3**: The causal diagram for the multiplier effects.

**Activity 3 - Tax multiplier**: We did not need to create a new causal diagram for Activity 3. The same causal diagram in Figure 3 is the basis of the tax multiplier activity because the same key variables and relationships are responsible for both the tax multiplier and the government-purchases multiplier. For this activity, we use the diagram to explain how changes in income tax, T, propagate through the system.

As Step 2 of the study, we prepared structural debriefing materials, including lecture slides and assignments. All activity forms, as distributed to students, can be found in the appendices at the end of this article.

## Implementation of the curriculum

In this section, we describe the implementation of debriefing activities during an Intermediate Macroeconomics course taught in Fall 2024 at a large research university by one of the authors. The class included undergraduate and graduate economics students. The graduate students were part of a new accelerated Master's track. The textbook used was Mankiw's "Macroeconomics" (2016). Activity 1 was implemented during a regularly scheduled 75-minute class. Activities 2 and 3 were combined and implemented during another 75-minute class. Before each activity, relevant textbook chapters were covered in a traditional manner, including lectures and homework assignments.

### Activity 1: National income model

This activity was introduced in the second week of the Intermediate Macroeconomics course, after the instructor had delivered regular lectures on the national income model and students had completed assigned homework. During the activity, the instructor used slides that familiarized students with the causal notation, walked them through the variables and causal relationships in the causal diagram, and provided examples of using the causal diagram for answering questions. We also developed a worksheet with three multiple-choice questions that students answered before and after the causal diagram was introduced. The expectation was that students would answer those questions more accurately with the help of the causal diagram. To simplify this task for students, we created a '*causal skeleton*' diagram in Figure 4. This diagram includes the same variables as in Figure 2 but uses directionless edges instead of signed arrows.

During class, the instructor followed this lesson plan:

1. Start the class by asking students to answer three multiple-choice questions in teams (see Appendix A). They must write their team # and student names. Collect their answers.
2. Open the slide deck for the activity. Explain the causal diagram notation, including arrow direction and polarity. Stop reviewing slides at this point -- do not explain slides beyond the notation.
3. Distribute the causal skeleton (Figure 4). Ask them to identify causal directions and polarity. Collect the assignment.
4. Go over the section of the slide deck that explains how to build the causal diagram for the national income model (Figure 2). A slide in Figure 5 shows an example of how the instructor would explain each causal link in the diagram, step-by-step. Note that explanations are numbered in the logical order -- 1, 2, 3, etc. Each text box corresponds closely to the narrative provided in the Mankiw textbook.
5. Distribute a completed causal diagram (Figure 2) as a solution to the exercise.
6. Explain how to use the causal diagram for answering questions. A slide with an example (Figure 6) asks what happens to the interest rate when government purchases decrease, shown on the graph as a red arrow pointing down. A negative causal arrow between government purchases and public savings suggests that public savings would increase, which is represented by a red arrow pointing up. If public savings increase, so do national savings. Then the instructor would continue following the causal chain until the answer is reached – the interest rate would decline.

7. Show the slide with Question 1 (Figure 7) from the multiple-choice set (Appendix A) without showing the solution. Invite students to answer it while using the causal diagram (Figure 2) that you just distributed. Collect their answers. Reveal the solution on the slide. Note that the slide includes a set of animated red arrows that show how an improvement in technology propagates through the system, leading to an answer.
8. Show students a slide with Question 2 and invite them to answer it. Encourage students to use the causal diagram. Ask them to write on a sheet of paper their answer. Collect their answers. Show students the solution to Question 2. Note that the slide provided to the instructor included an animation that showed with red arrows how an improvement in technology propagates through the system, leading to an answer.
9. Show students a slide with Question 3 and invite them to answer it. Encourage students to use the causal diagram. Ask them to write on a sheet of paper their answer. Collect their answers. Show students the solution to Question 3. Note that the slide includes an animation that shows how a change propagates through the system, leading to an answer.
10. Distribute a questionnaire.

For this activity, the class of 15 students was divided into five teams of three students each. The instructor followed the lesson plan and used a set of slides that guided the activity. Each group received a worksheet with three multiple-choice questions (see Appendix A). After recording their answers, the teams returned the worksheets to the instructor. These questions with correct answers in bold are presented in Appendix A. After the instructor collected the answers to the three questions, the causal diagram of the national income model was explained. Then students were given the opportunity to answer the three questions again, this time using the causal diagram. The instructor concluded the class by inviting students to complete a feedback questionnaire.

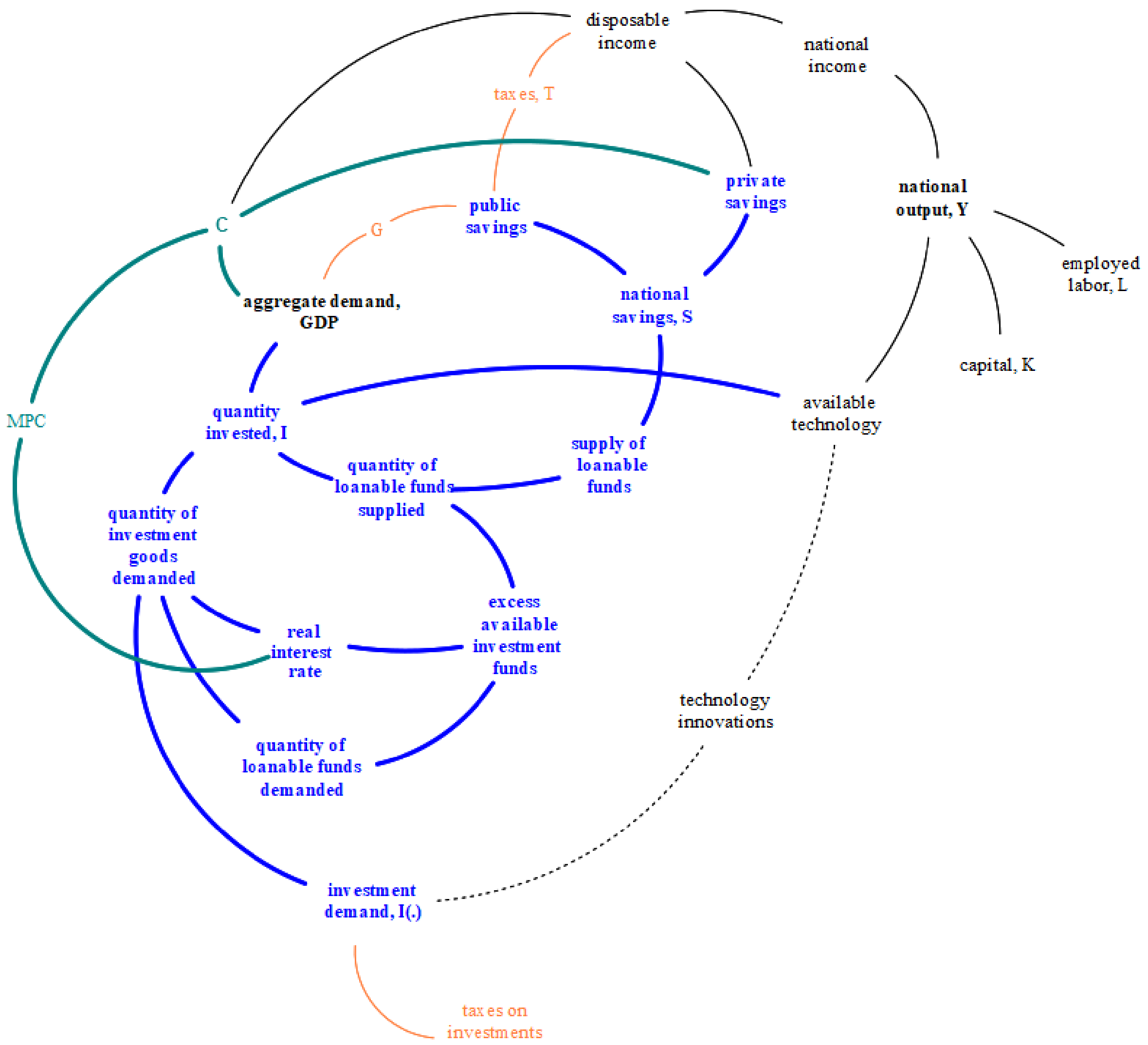

**Figure 4**: A causal skeleton of the National Income Model

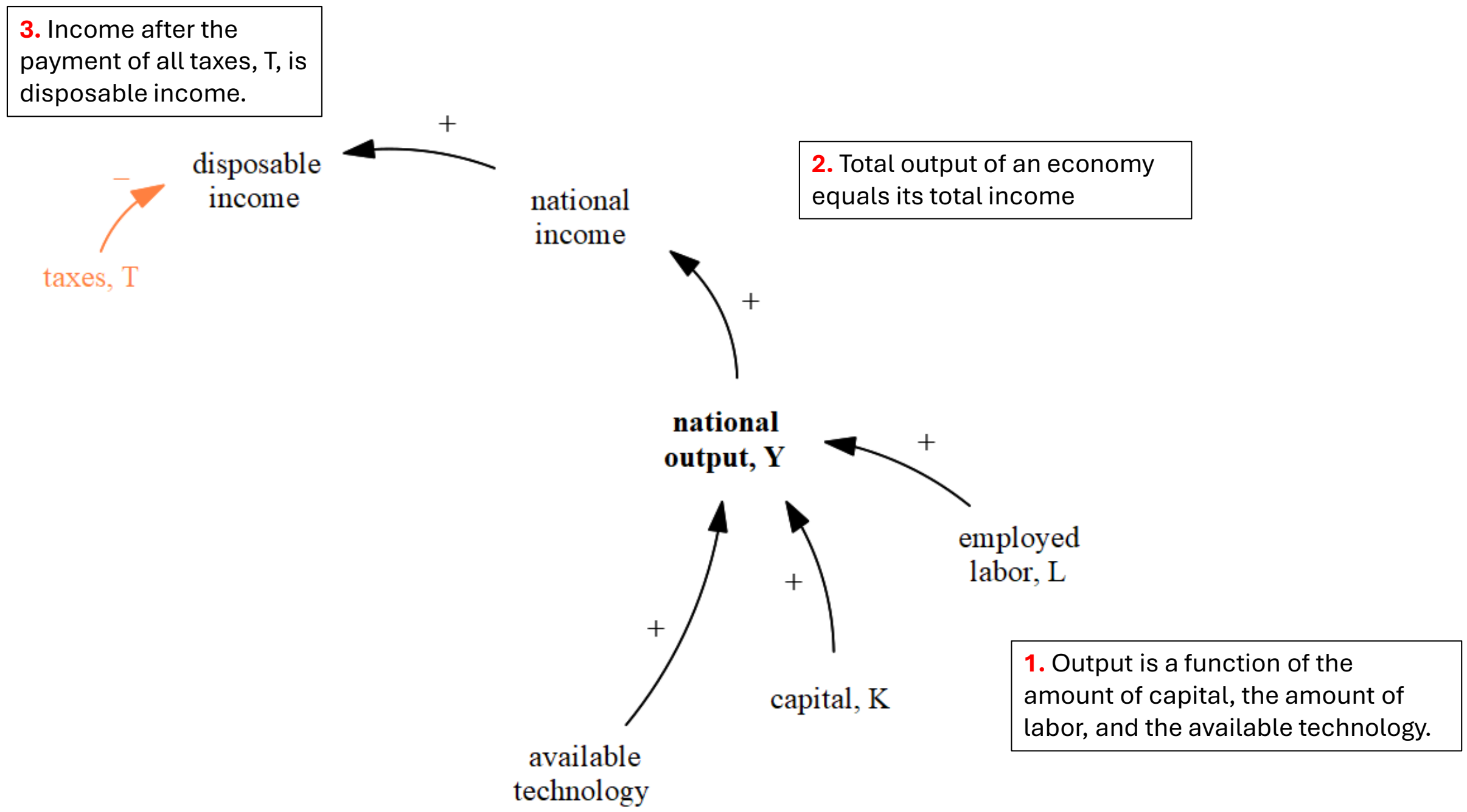

**Figure 5**: A slide that explains a portion of the causal diagram for the national income model.

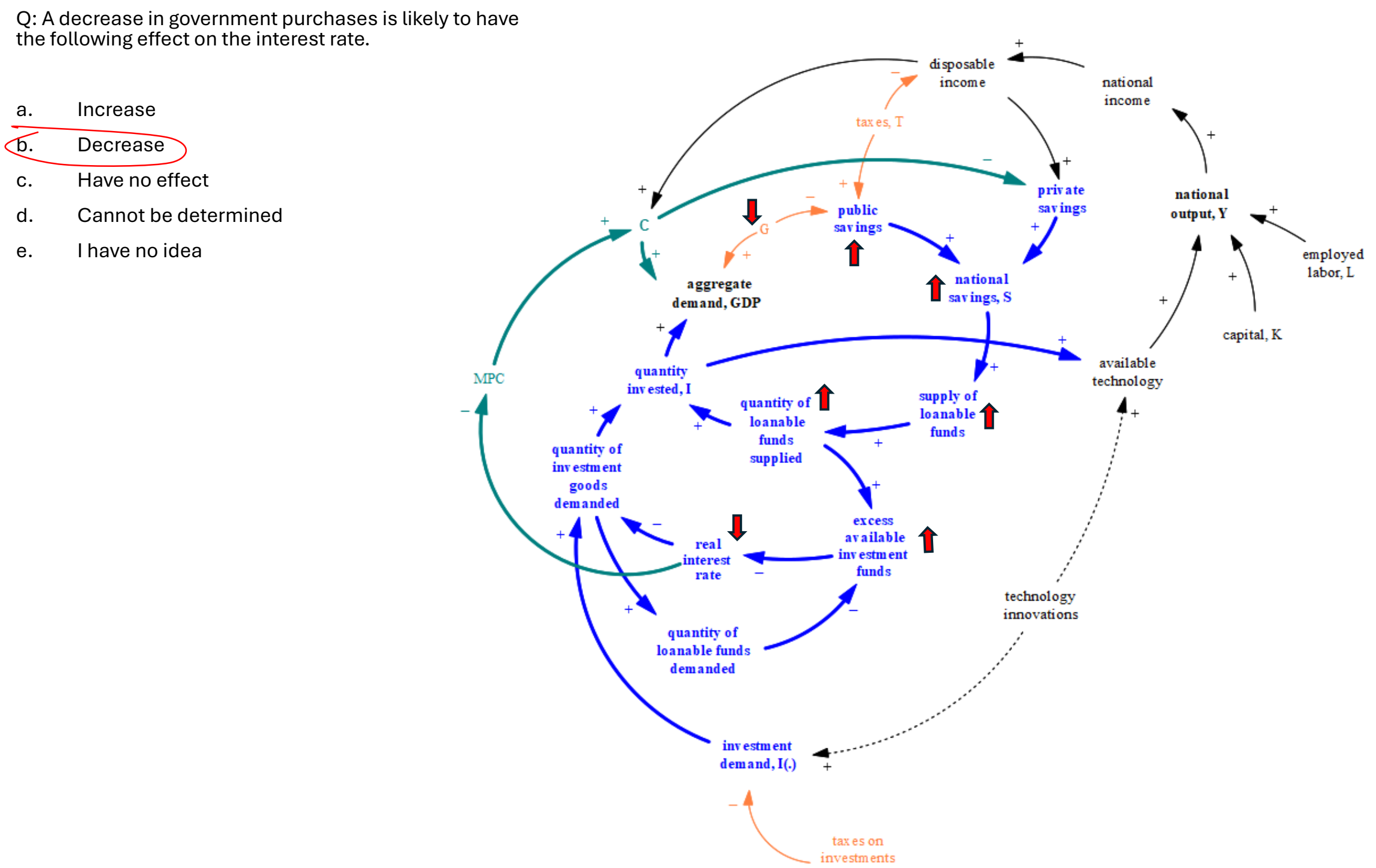

**Figure 6**: A slide that shows how to answer a multiple-choice question using the causal diagram of the national income model. The question asks what would happen to the interest rate when government purchases decrease.

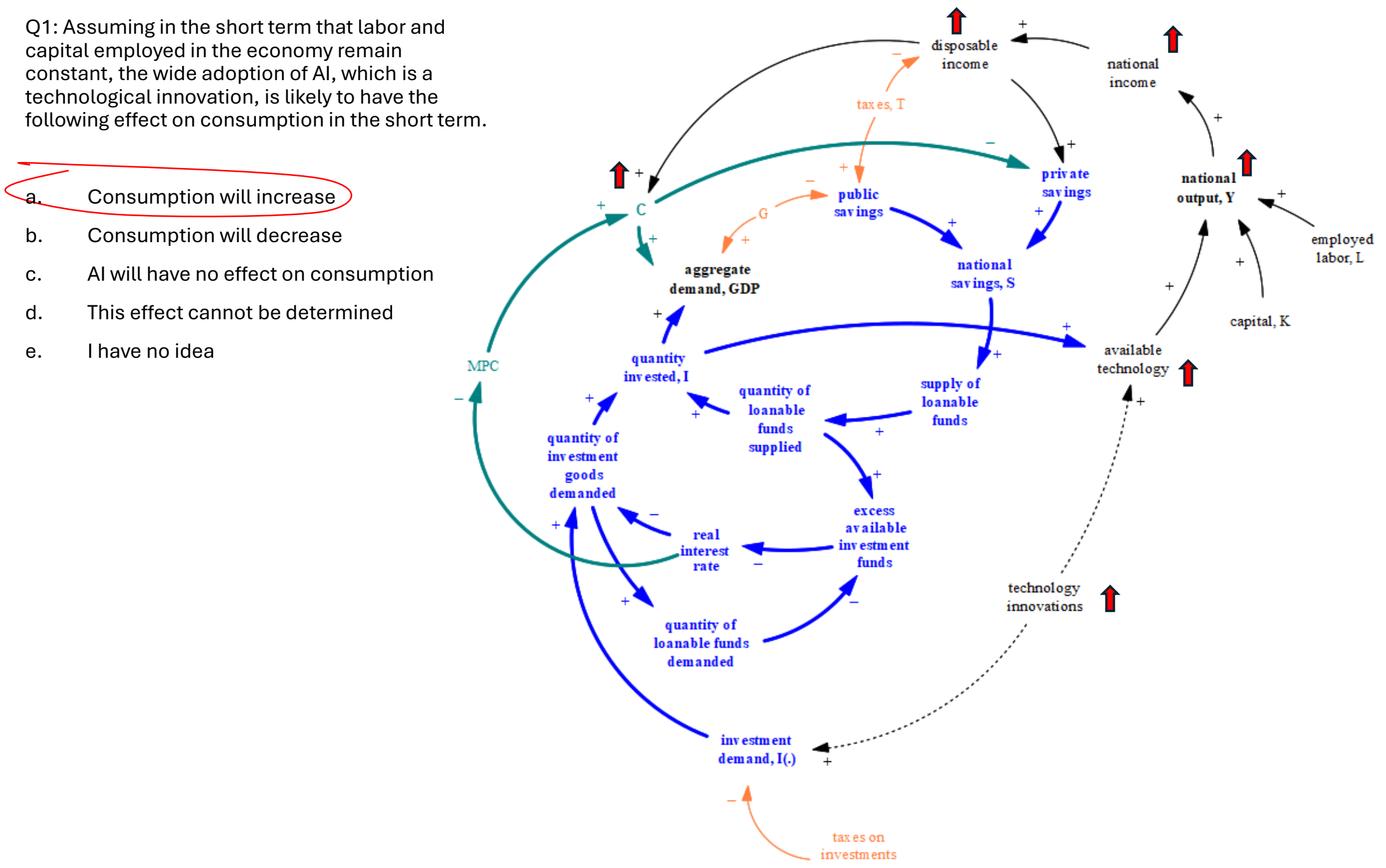

**Figure 7**: A slide that shows how to answer a multiple-choice question using the causal diagram of the National Income Model.

**Activity 2: Government-purchases multiplier**

Incorporating the experience of Activity 1, we prepared a set of slides and an Instructor Manual for Activities 2 and 3, which were conducted during the same class. Appendices B, C, and D provide the materials used for the multiplier activities. Appendix E includes the Instructor Manual.

Thirteen students participated in Activity 2. They worked individually. While students had already covered the material on the government-purchases multiplier in earlier classes, we knew from our in-class observations during Activity 1 that students might not be able to recall easily relevant knowledge. Therefore, for Activity 2 we quoted applicable passages from the textbook, which students had seen before. Students received hard copies of Form 1 (see Appendix B) that asked them to complete a causal skeleton graph (Figure 8) using the provided quotes. They had 10 minutes for this task.

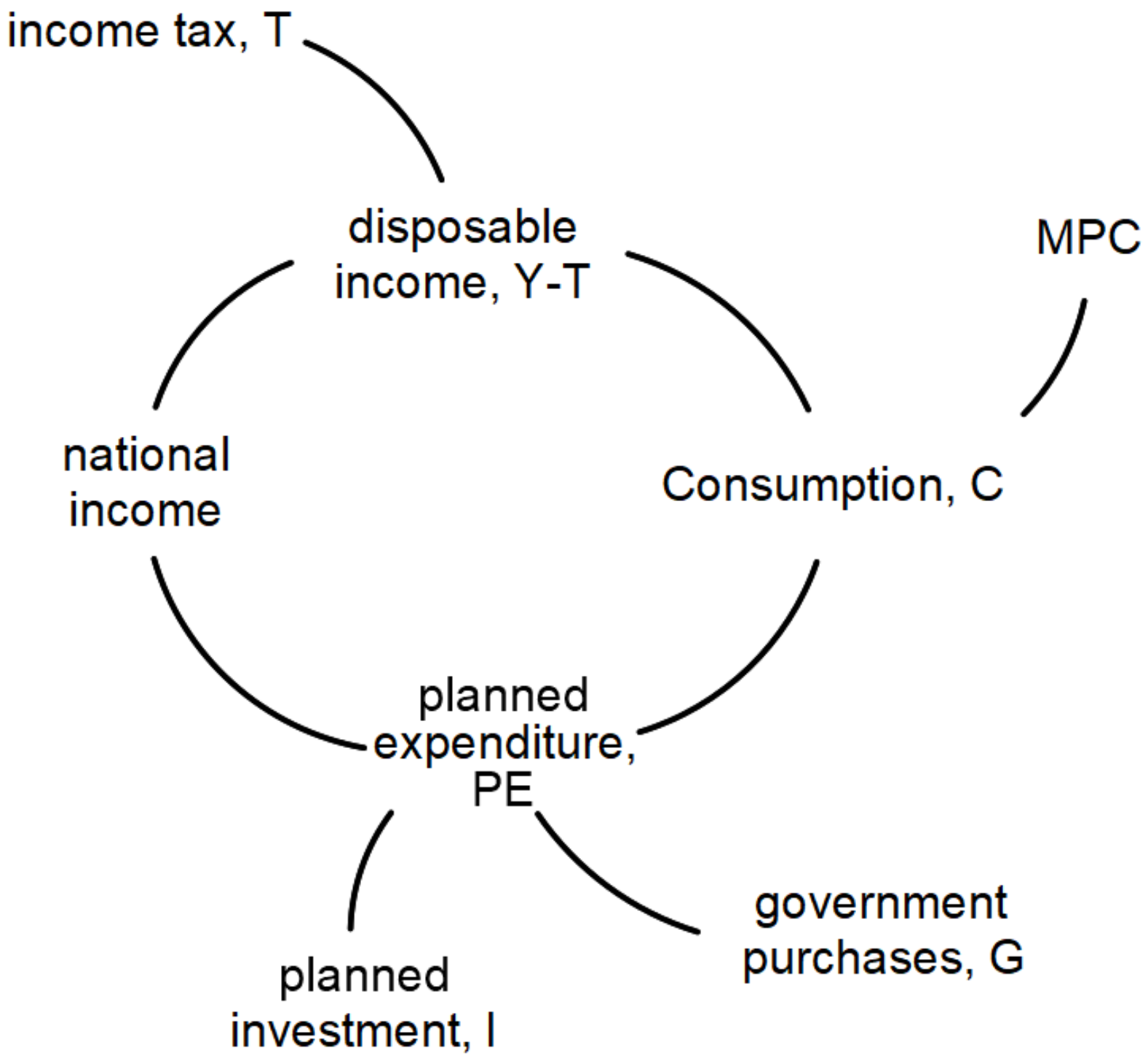

**Figure 8:** A causal skeleton graph given to students during Activities 1 and 2

After submitting their answers, students were shown the finished causal diagram. The instructor also used a prepared slide deck to explain step-by-step how an initial increase in government purchases, $\Delta G$, would propagate multiple times through the feedback loop resulting in the multiplier effect. Figure 9 shows the corresponding graphic. When government purchases increase by $\Delta G$, planned expenditure, national income and disposable income rise by $\Delta G$ as well. But consumption increases only by $(MPC \times \Delta G) < \Delta G$ because $MPC < 1$. This is the first change in consumption due to $\Delta G$. This change in consumption increases planned expenditure by $MPC \times \Delta G$, which starts the second iteration around the feedback loop. Note that the second increase in planned expenditure due to $\Delta G$ is less than the first increase. The numbers above the red arrows correspond to the first, second, and third iterations around the feedback loop. The lengths of the red arrows signify the relative scale of the effects from the initial $\Delta G$.

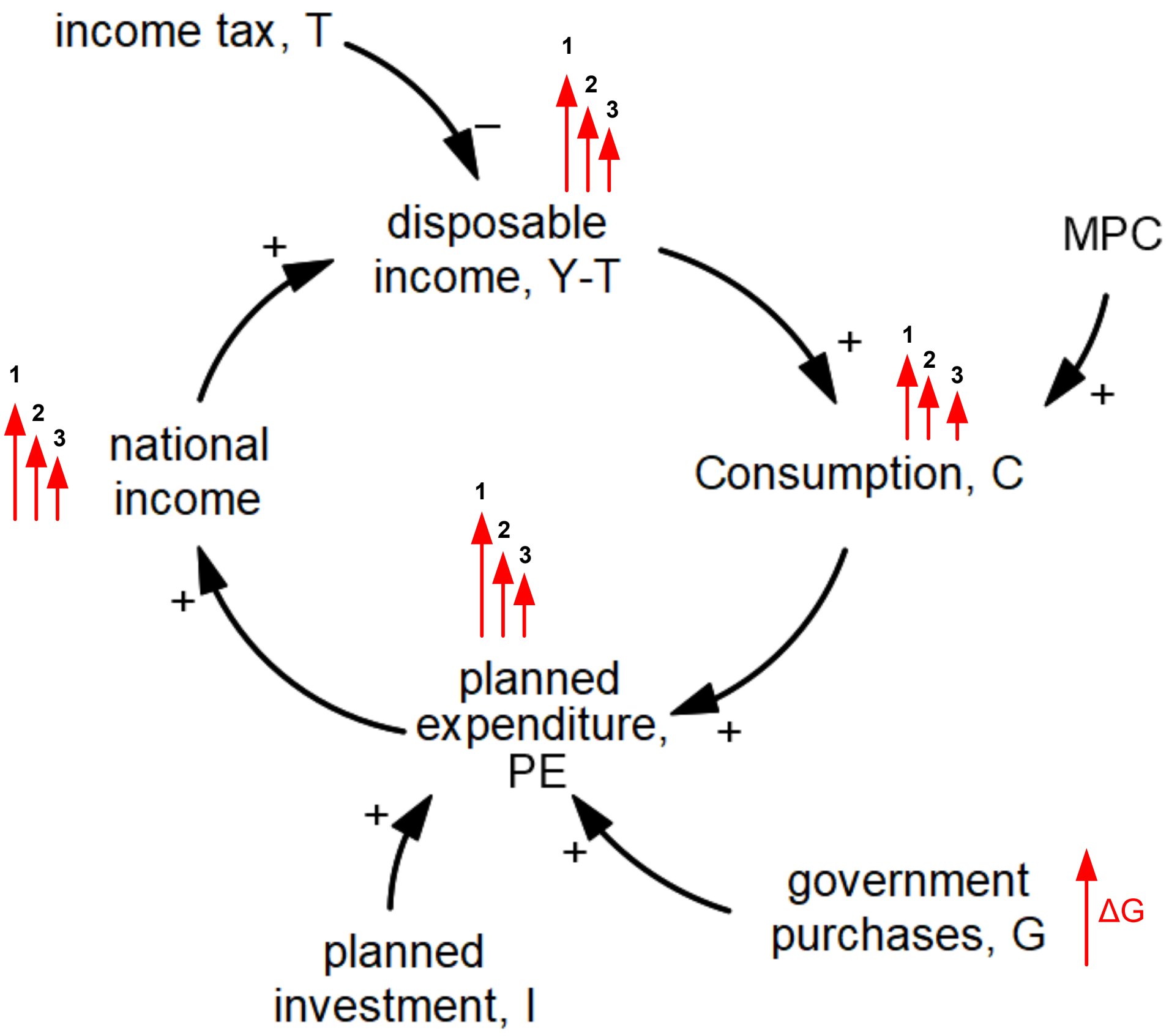

**Figure 9:** The government-purchases multiplier effect.

Corroborating textbook explanations, after completing three iterations around the loop, the instructor pointed out that the total effect from *ΔG* on national income is:

| | |
|---|---|
| Initial Change in Government Purchases | $= \Delta G$ |
| First Change in Consumption | $= MPC \times \Delta G$ |
| Second Change in Consumption | $= MPC^2 \times \Delta G$ |
| Third Change in Consumption | $= MPC^3 \times \Delta G$ |
| ⋮ | ⋮ |

$\Delta Y = (1 + MPC + MPC^2 + MPC^3 + \ldots)\ \Delta G$

**Activity 3: Tax multiplier**

One student was late to class, and therefore missed Activity 2. This student, however, took part in Activity 3, and therefore 14 students participated in the last activity.

This activity was similar to Activity 2. Students received a worksheet (see Form 2 in Appendix C) that quoted passages from Mankiw's "Macroeconomics" about the tax multiplier effect. The form also included a causal skeleton as in Figure 8 that students were asked to complete by adding causal directions and link polarities. This was their second time completing the same causal skeleton on that day.

Then, we asked students to explain the tax multiplier effect using the causal graph that they just completed. Students had seen this type of analysis in Activity 2, when the instructor explained how a one-time increase in government purchases adds multiple times to consumption. After submitting their worksheets, students were shown the causal diagram that included the analysis of a tax decrease, ΔT (see Figure 10).

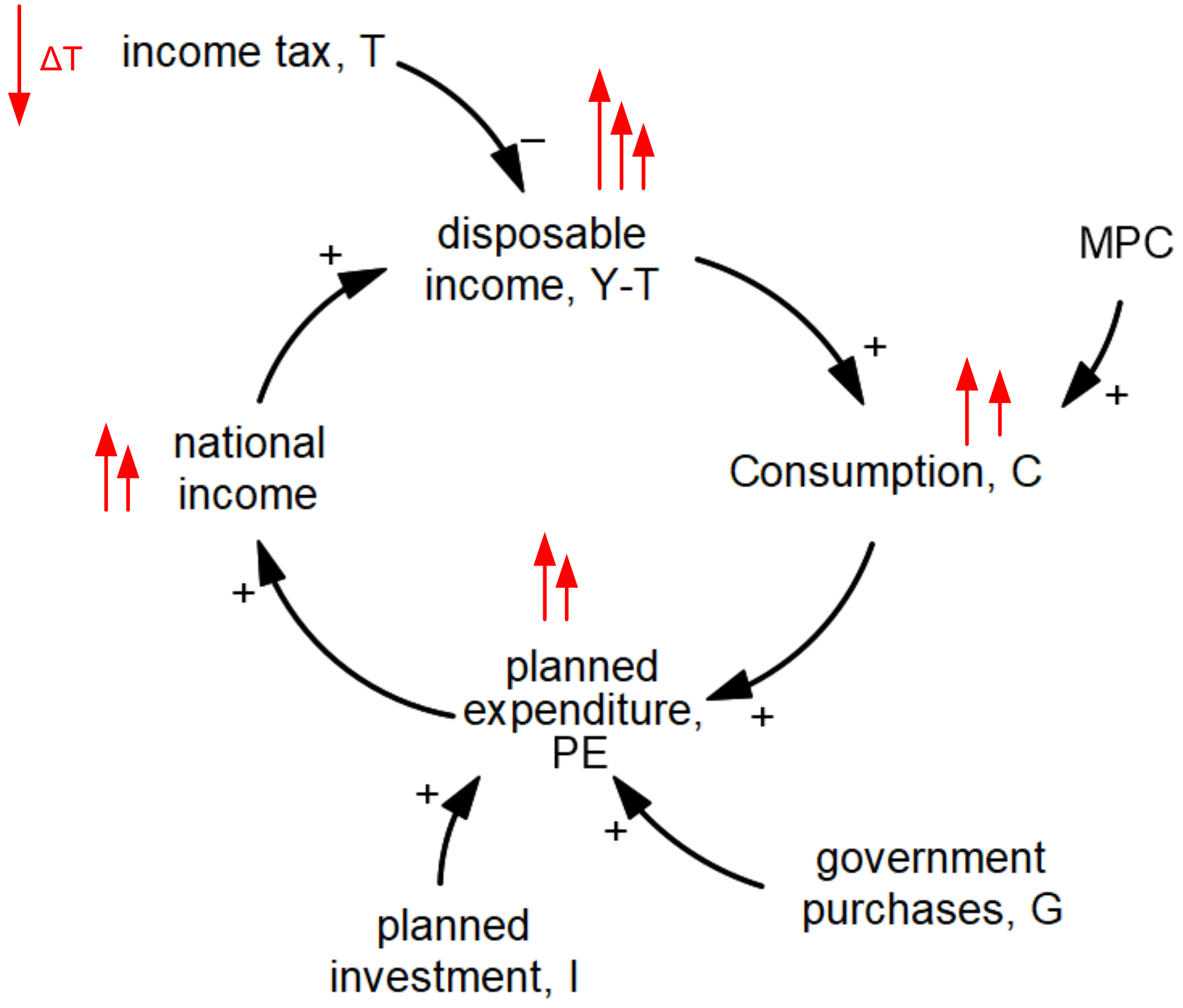

**Figure 10**: The tax multiplier effect.

The instructor also noted that the total effect on national income is:

| | |
|---|---|
| Initial Change in Taxes | $= - \Delta T$ |
| First Change in Consumption | $= MPC \times \Delta T$ |
| Second Change in Consumption | $= MPC^2 \times \Delta T$ |
| Third Change in Consumption | $= MPC^3 \times \Delta T$ |
| ⋮ | ⋮ |

$$\Delta Y = (1 + MPC + MPC^2 + MPC^3 + \ldots)\ \Delta T$$

These calculations can also be found in Mankiw (2016). Note that because taxes are reduced, the initial change in taxes is negative, $-\Delta T$. However, due to the inverse causal relationship between taxes, T, and disposable income, $Y - T$, the change in disposable income is positive (it is equal to $\Delta T$), and the first change in consumption is $MPC \times \Delta T$.

After completing Activities 2 and 3, students were invited to answer five questions (see Form 3 in Appendix D). Two of these questions required students to discuss similarities and differences between the government-purchases and tax multipliers, both of which share the same causal structure depicted in Figure 4. These comparison questions were inspired by Goffe and Wolla's (2024: 159) suggestion to use deliberate practice for developing schemas in an economics course, such as by asking students to consider how income and price elasticities are similar and different.

## Findings

This section analyzes the implementation of the debriefing activities.

### Activity 1: National income model

Students completed the causal skeleton worksheet (Figure 4) by drawing the arrow directions and marking the polarity of the relationships -- a plus sign for a positive effect and a minus sign for a negative effect. Table 1 documents student responses. The diagram has 32 arrows, and therefore 32 correct answers for both direction and polarity are possible. As seen from the table, Student 7 provided the most accurate responses by identifying correctly 53% of causal directions and 69% of causal polarities. The class averages were 24% and 37% for directions and polarities respectively. The relatively poor performance suggests that even after covering the national income model in class, students either struggled to recall the textbook material or they were unclear about the causal notation and working with causal diagrams. A multiple-choice exercise that we report on next suggests that students were most likely uncertain on how to use the causal diagram.

**Table 1.** Student responses to the national income model causal direction and sign exercise

| | Correct answers | | Percentage correct | |
|---|---|---|---|---|
| Student | Direction | Polarity | Direction | Polarity |
| 1 | 0 | 15 | 0.00% | 46.88% |
| 2 | 0 | 0 | 0.00% | 0.00% |
| 3 | 3 | 3 | 9.38% | 9.38% |
| 4 | 4 | 2 | 12.50% | 6.25% |
| 5 | 0 | 13 | 0.00% | 40.63% |
| 6 | 0 | 14 | 0.00% | 43.75% |
| 7 | 17 | 22 | 53.13% | 68.75% |
| 8 | 13 | 19 | 40.63% | 59.38% |
| 9 | 16 | 21 | 50.00% | 65.63% |
| 10 | 4 | 14 | 12.50% | 43.75% |
| 11 | 6 | 11 | 18.75% | 34.38% |
| 12 | 16 | 17 | 50.00% | 53.13% |
| 13 | 15 | 13 | 46.88% | 40.63% |
| 14 | 9 | 8 | 28.13% | 25.00% |
| 15 | 10 | 6 | 31.25% | 18.75% |
| Mean | 7.53 | 11.87 | 23.54% | 37.08% |
| Median | 6.00 | 13.00 | 18.75% | 40.63% |
| St Deviation | 6.56 | 6.82 | 20.49% | 21.32% |

Students were invited to answer the same three questions (Appendix A) before and after the causal diagram exercise. They worked in five teams. Their answers are recorded in Table 2 as pre- and post-diagram results. Each team answered three questions, and therefore the total number of possible correct answers by five teams was 15. We can see that pre-diagram three teams answered all questions correctly. In total, the number of correct pre-diagram answers was 12 out of 15, or an 80% correct response rate. When students had access to the solution causal diagram (as in Figure 2), only one team improved its results -- see the post-diagram numbers and the last column in Table 2. Two teams did worse, and two teams performed the same. Overall, post-diagram, there were 10 correct responses, or 67% correct response rate. This drop in the overall performance on multiple-choice questions was unexpected as these were the same questions that students had answered before. We attribute this overall drop in performance to student confusion regarding how to use the causal diagram.

**Table 2.** Pre- and post- National Income causal diagram exercise results.

| | Correct answers | | | | |
|---|---|---|---|---|---|
| | Pre-diagram | | Post-diagram | | |
| | Count | % | Count | % | Result |
| Team 1 | 3 | 100 | 2 | 67 | Worse |
| Team 2 | 3 | 100 | 1 | 33 | Worse |
| Team 3 | 3 | 100 | 3 | 100 | Same |
| Team 4 | 1 | 33 | 2 | 67 | Better |
| Team 5 | 2 | 67 | 2 | 67 | Same |
| Total | 12 | 80 | 10 | 67 | Worse |

*Student feedback to Activity 1*

Students were invited to provide their feedback for the activity, including their perception of positive experience and constructive criticism. Students responded to the following prompts:

**Q1:** Was the causal diagram helpful for deeper understanding of the national income model? (YES/NO)
**Q2:** List three things that you have learned from the causal diagram of the national income model.
**Q3:** Any other comments?

In response to the question if the causal diagram was helpful (Q1), 10 students out of 15 answered positively, two students said that the diagram was not helpful, and three students did not answer that question.

When listing three things that they learned from the causal diagram of the national income model (Q2), those who found the causal diagram helpful wrote:

a. *"There are many unseen implications of the National Income Model"*
b. *"That an increase in Public Savings leads to lower interest rate"*
c. *"That technical innovation leads to an increase in quantity invested"*

The same student added in response to Q3*: Thank you very much for this!*

a. *"How things are interrelated"*
b. *"Directions of flows for each process"*
c. *"Systems of equations are easier"*

a. *"The direction of relationships"*
b. *"There isn't a starting point"*
c. *Y+/- ΔI and similar relationships within National Income model*

The same student commented in response to Q3*: The chart is helpful. Maybe add more equations below referencing variables.*

*a. "Relationship between public, private and national savings"*
*b. "How interest rates affect Government spending"*
*c. "Disposable income takes part in the market for loanable funds"*
A Q3 comment*: I'd like this activity for future chapters.*

*a. "Decrease in government purchases will have a decrease in interest rate"*
*b. "The more national savings, the more supply of loanable funds"*
c. *"The more quantity invested, the higher the GDP"*

The two students who thought that the causal diagram was not helpful reported these takeaways, including the three things that they learned:

*a. "The relationship of government purchase and interest rate"*
*b. "The flows of money"*
*c. "Economics is cool"*
Comment*: This is too much info at once for me to efficiently learn*

*a. "If disposable income is higher, taxes go up. And if taxes go up, disposable income goes down"*
*b. "Desire to invest increases loanable funds demanded"*
*c. "Excess funds reduce interest rates"*
Comment*: It would be helpful to separate the colors and show it (National Income Model) by section.*

Overall, the students enjoyed the classroom exercise and were actively engaged.

**Activity 2: Government-purchases multiplier**

Table 3 presents the results from a "Government-Purchases Multiplier" exercise, when students were asked to complete a causal skeleton by adding arrow directions and polarities. Thirteen students completed the worksheet (see Appendix B). The causal diagram had eight variables and eight edges.

**Table 3:** Results of the activity to complete a causal diagram for the government-purchases multiplier (Form 1)

| Student | Correct answers | | Percentage correct | |
|---|---|---|---|---|
| | Direction | Polarity | Direction | Polarity |
| 1 | 8 | 6 | 100.00% | 75.00% |
| 2 | 7 | 0 | 87.50% | 0.00% |
| 3 | 7 | 6 | 87.50% | 75.00% |
| 4 | 7 | 0 | 87.50% | 0.00% |
| 5 | 0 | 3 | 0.00% | 37.50% |
| 6 | 7 | 7 | 87.50% | 87.50% |
| 7 | 5 | 8 | 62.50% | 100.00% |
| 8 | 5 | 8 | 62.50% | 100.00% |
| 9 | 6 | 8 | 75.00% | 100.00% |
| 10 | 8 | 6 | 100.00% | 75.00% |
| 11 | 4 | 1 | 50.00% | 12.50% |
| 12 | 6 | 7 | 75.00% | 87.50% |
| 13 | 3 | 7 | 37.50% | 87.50% |
| Mean | 5.62 | 5.15 | 70.19% | 64.42% |
| Median | 6.00 | 6.00 | 75.00% | 75.00% |
| St Deviation | 2.26 | 3.05 | 28.20% | 38.14% |

The results show that on average 70% of students correctly drew the arrow directions, while 64% correctly added the arrow polarities. The median for correct answers was 75% for both arrow directions and polarities. The standard deviations were 28% and 38%, respectively.

### Activity 3: Tax multiplier

Students completed the worksheet for Activity 3 (see Appendix C) immediately after they finished Activity 2 and were presented with the solution for the government-purchases multiplier diagram. Table 4 tabulates the results for Activity 3. While fourteen students participated in the activity, interestingly, one student did not answer any questions. There are several students who marked all arrow directions and polarities correctly. As a result, the averages improved relative to the similar government-purchases multiplier activity – this time, 83% of answers were correct for arrow directions and nearly 70% for arrow polarities.

**Table 4:** Results of completing the causal diagram for the tax multiplier (Form 2)

| | Correct answers | | Percentage correct | |
|---|---|---|---|---|
| Student | Direction | Polarity | Direction | Polarity |
| 1 | 8 | 7 | 100.00% | 87.50% |
| 2 | 8 | 8 | 100.00% | 100.00% |
| 3 | 4 | 0 | 50.00% | 0.00% |
| 4 | 0 | 0 | 0.00% | 0.00% |
| 5 | 8 | 7 | 100.00% | 87.50% |
| 6 | 5 | 7 | 62.50% | 87.50% |
| 7 | 8 | 8 | 100.00% | 100.00% |
| 8 | 8 | 7 | 100.00% | 87.50% |
| 9 | 8 | 5 | 100.00% | 62.50% |
| 10 | 8 | 8 | 100.00% | 100.00% |
| 11 | 5 | 0 | 62.50% | 0.00% |
| 12 | 7 | 5 | 87.50% | 62.50% |
| 13 | 8 | 8 | 100.00% | 100.00% |
| 14 | 8 | 8 | 100.00% | 100.00% |
| Mean | 7.15 | 6.00 | 83.04% | 69.64% |
| Median | 8.00 | 7.00 | 100.00% | 87.50% |
| St Deviation | 1.46 | 2.86 | 29.66% | 39.75% |

*Student feedback to Activities 2 and 3*

After completing Activities 2 and 3, students were invited to fill out a questionnaire that included five questions (see Appendix D).

Question 1 probed students' understanding that the same economic mechanisms are responsible for the government-purchases multiplier and the tax multiplier. Students recognized similarities between the multipliers. Here are some examples of their responses:

> *"Both government purchases multiplier and tax multiplier affect national income through changes in planned expenditure. They both influence disposable income."*

> *"Yes, both increase consumption, national income, PE, and disposable income. They both increase economic growth."*

> *"Yes, both the government purchases multiplier and the tax multiplier are based on the same core economic principles, where the initial change in spending leads to instability in the economy."*

> *"Yes, same variables are involved, and they have the same causal relationships."*

Question 2 was, “Are there any differences between the economic structures that are responsible for the government purchases multiplier and the tax multiplier?” Some students pointed out that government purchases might have a stronger impact dollar-for-dollar on the GDP due to its direct impact while a tax reduction acts on GDP through consumption. Here are examples of student answers:

> *“Government purchases multiplier typically has a larger impact compared to tax multiplier due to direct increase in planned expenditure.”*

> *“Government purchases multiplier directly impacts GDP at a rapid way, and not all tax multiplier goes into the GDP, as some is saved which has weaker effect”.*

> *“Government purchases multiplier has a more direct impact on aggregate demand, while the tax multiplier works indirectly by increasing disposable income”.*

Then we asked students if the causal diagrams helped them understand the two multiplier effects (Question 3), or whether any part of the activities was confusing (Question 4). The class expressed a range of opinions on these related questions. Below are some answers we received:

> *“Yes, I like the visual diagram to see the flow of impact.”*

> *“Yes, they [causal diagrams] are helpful. They visually explained the flow of causal relationships and feedback loops and make it easier to see how a change in one variable propagates through the economy.”*

> *“No, they [causal diagrams] confused me. I failed to understand the direction of the arrows, although I read the textbook.”*

> *“It is still a little confusing. The arrows and the signs sometimes didn’t add up.”*

> *“I think the formulas were more helpful, causal diagrams stopped helping after I understood the relationship between all of the variables (taxes and disposable income, for example)...Causal diagram was not confusing, but on Form 2 when it asked to show the tax multiplier, that was confusing...I knew things like disposable income would increase thanks to the formula but past that I didn’t know how else to show it.”*

> *“Yes, I like the layers as the arrows, signs, and arrows help show exact effects...I think the confusing part was the differentiation between both at first.”*

> *“The slides [gave] the same information and it was clear enough / displayed the animation better than pencil/paper...Mankiw states “when an increase in government purchases raises income, it also raises consumption...” Where this diagram only flows [in] one direction, Mankiw doesn’t explicitly describe the directional flow.”*

*"They [causal diagrams] were helpful in a closed model, but would it still make sense in an open economy? I am still confused a bit on which way the arrows are supposed to point. Do they point to the thing they caused or are causing?"*

*"Yes. They were helpful in seeing the effects visually."*

*"Yes it showed the dependency"*

*"Yes, [causal diagrams] show the direction of movement in the economic structure."*

Question 5 asked students to apply the causal diagram approach to explain situations that they face at work and home or read about in the news. Here are several responses:

*"The way I understand causal diagrams is that they just depict a relationship between two economic variables. One example I could give is*

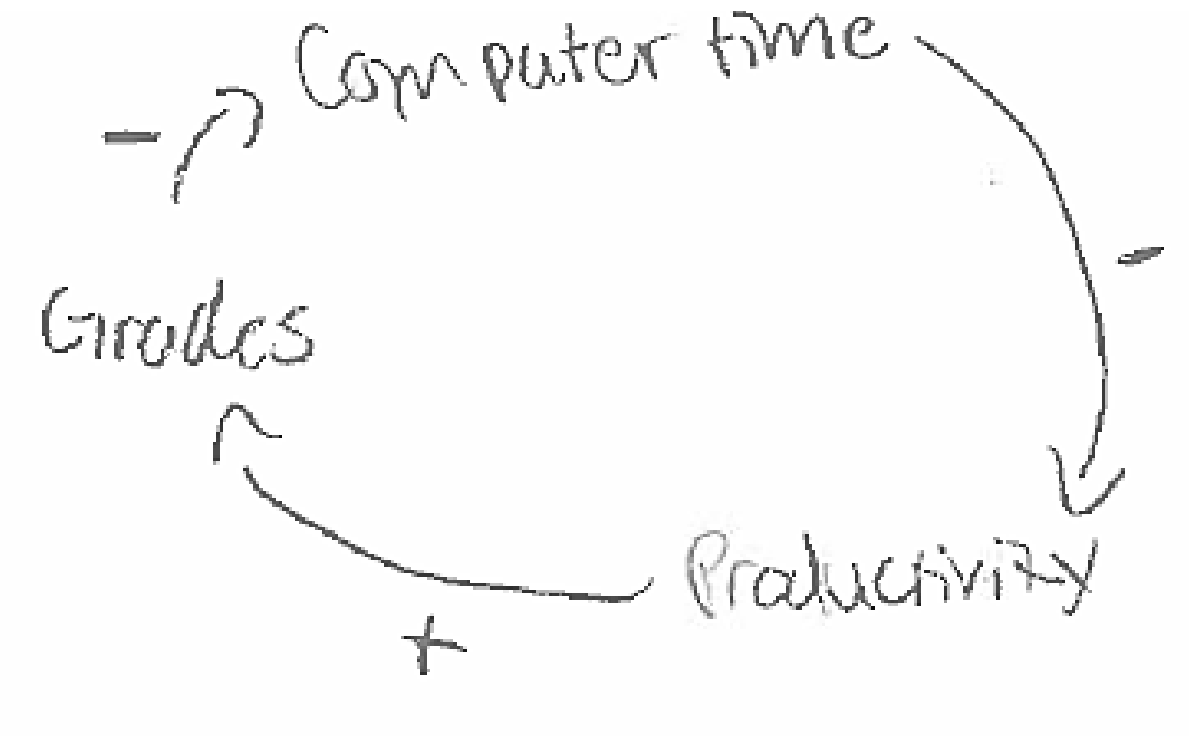

*"Yes, [causal diagrams] help trace "train of thoughts" while focusing on the causation."*

*"Yes, flowcharts are very helpful to describe many systems. These particular flowcharts are very theoretical, and I would be hesitant to consider their assumptions without empirical proof"*

*"I think it is possible to use them in other areas to diagram impact and see how one change impacts the cycle."*

*"To some extent. For example, see below for an example of someone who is working a job:*

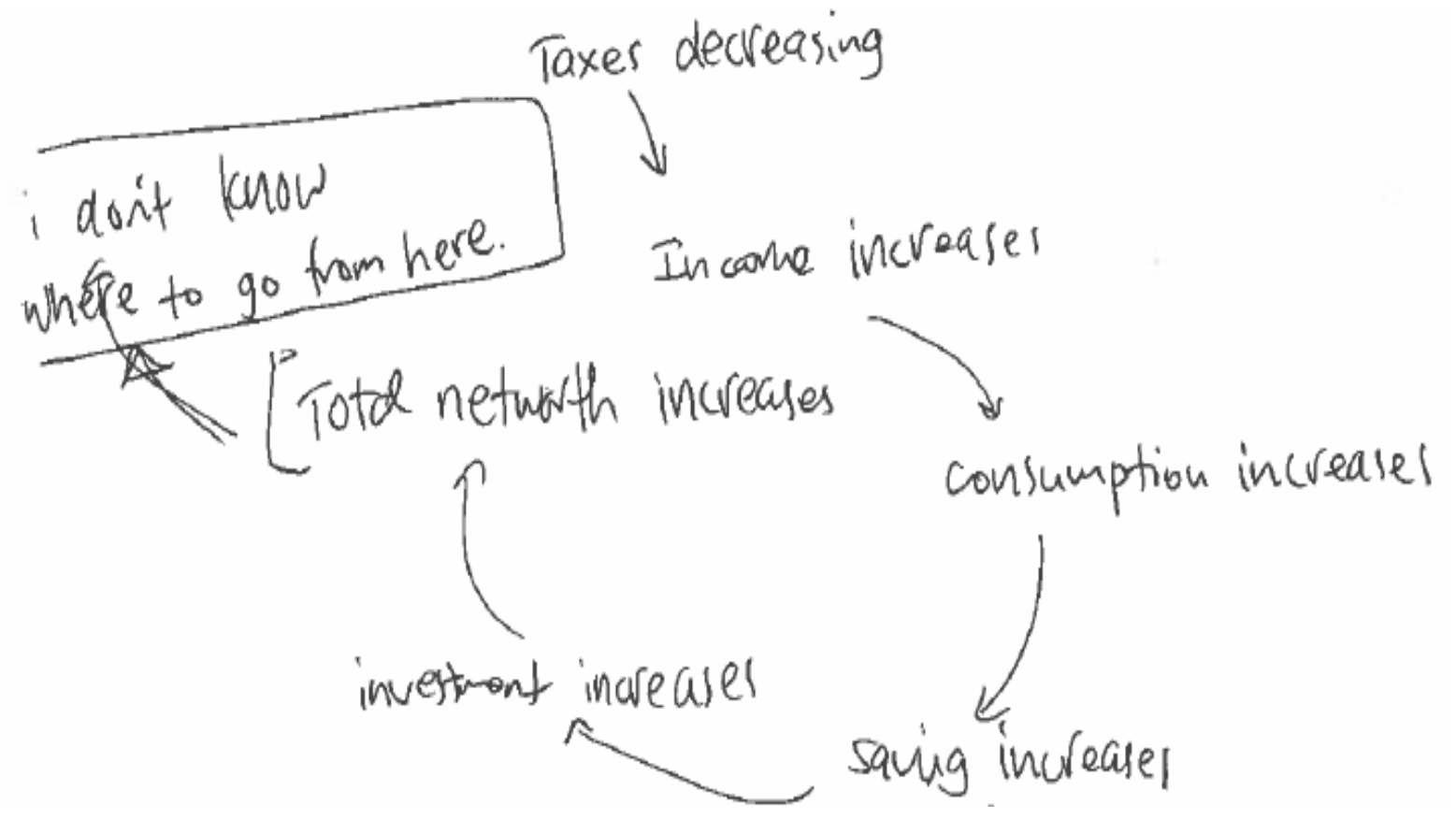

*"Yes. An example I can think about is my personal consumption and how that plays a part in our economy"*

*"Yes. Things like the effect of change in interest rates can be visualized with the causal diagram"*

*"Yes, why fiscal stimulus is used more often / is more effective in large change during emergency vs decreasing taxes"*

## Discussion

The objective of this study was to answer the research question of how instructors can utilize causal diagrams and structural debriefing activities for literacy-targeted instruction in a macroeconomics classroom. Three activities were developed and implemented.

The implementation of Activity 1 highlighted that reading and interpreting causal diagrams was a challenging task for students, a finding also noted by Capelo et al. (2024) in their structural debriefing study. Also, it became evident that students could not be expected to recall numerous variables and relationships from memory, even if the economic material had been covered in class and hints were provided in the form of a causal skeleton. For example, one student commented after Activity 1, "*This is too much info at once for me to efficiently learn.*" Consequently, Activities 2 and 3 were designed around smaller causal diagrams, and the worksheets included passages from the textbook to help students complete the causal skeletons.

For additional insights, we calculated class statistics for each activity and tracked the progress of the 10 students who participated in all three activities. These data are reported next.

### Class statistics

Table 5 summarizes class statistics for the three activities, which are also visualized in Figure 11. In addition to the arithmetic means and standard deviations (SD) from Tables 1, 3, and 4, we calculated the coefficient of variation (CV) for each activity. The coefficient of variation, defined as SD/Mean, measures the variability of results. During Activity 1, approximately 24% of arrow directions and 37% of arrow signs were identified correctly on average. This relatively poor outcome improved significantly after we made adjustments for the subsequent activities. In Activity 2, the mean for correct arrow directions increased to 70%, and the mean for correct arrow signs improved to 64%. For Activity 3, on average, 89% of causal directions and 75% of causal signs were accurately marked. This improvement in students' ability to identify cause-and-effect relationships is clearly illustrated in Figure 11a. While the standard deviations remained quite significant throughout the activities, the variability of results for direction identification declined with experience. This change is evident in the first CV column in Table 5 and in Figure 11b. However, Figure 11b and the second CV column suggest that some students continued to struggle when determining arrow polarity.

**Table 5:** Class statistics for correct answers

| | Direction | | | Polarity | | |
|---|---|---|---|---|---|---|
| | Mean | SD | CV | Mean | SD | CV |
| Activity 1 | 23.54% | 20.49% | 0.87 | 37.08% | 21.32% | 0.57 |
| Activity 2 | 70.19% | 28.20% | 0.40 | 64.42% | 38.14% | 0.59 |
| Activity 3 | 83.04% | 29.66% | 0.36 | 69.64% | 39.75% | 0.57 |

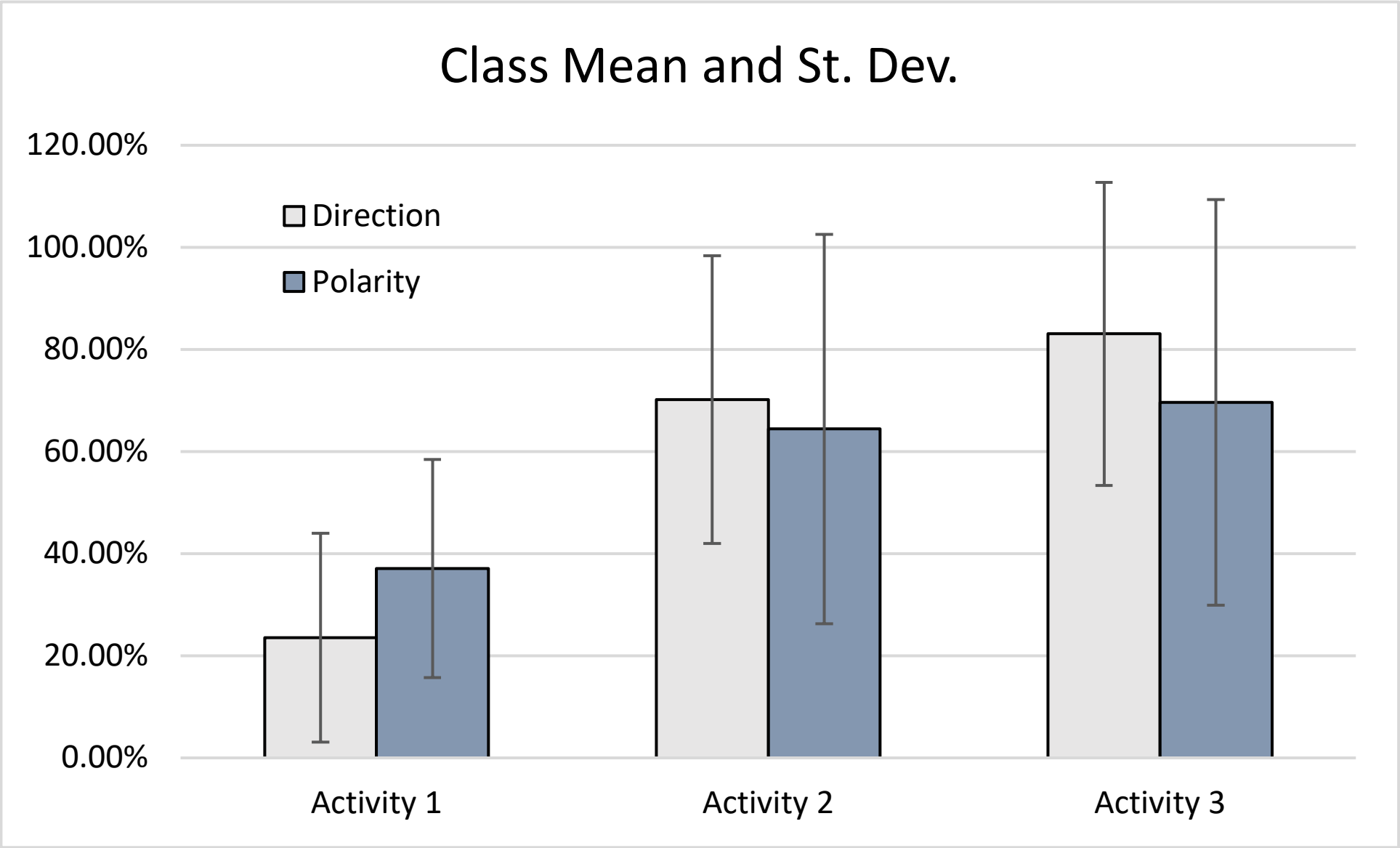

(a)

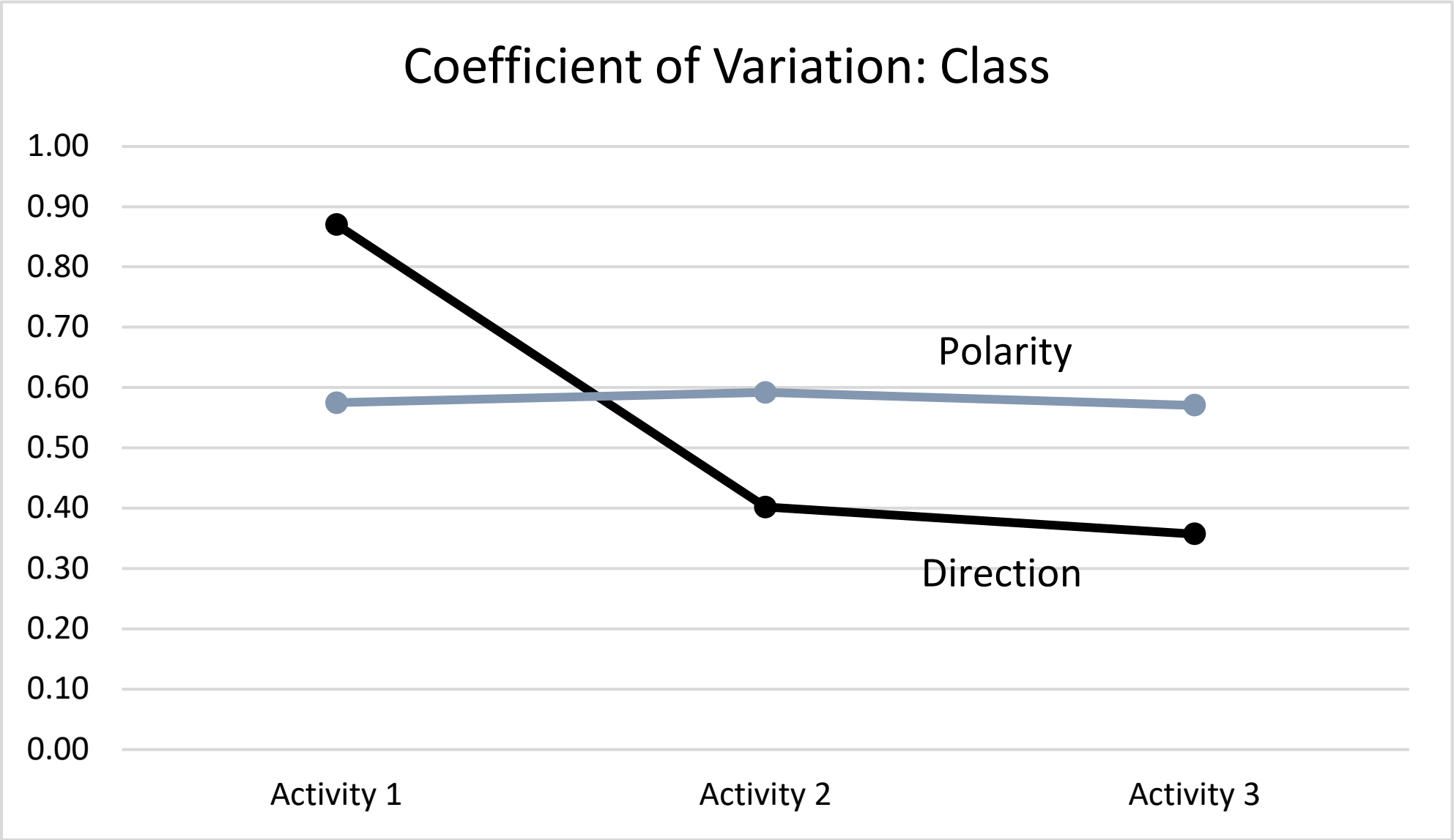

(b)

**Figure 11:** Class statistics for correct answers during the activities.

**Statistics for 10 persistent students**

In this section, we review the performance of the 10 students who participated in all three activities. In Table 6 we compiled their outcomes for each activity that are also presented graphically in

Figure 12. Figure 12a shows that all 10 students improved their performance in direction identification from Activity 1 to Activity 2. This trend continued into Activity 3, with the exception of two students who performed worse in Activity 3.

**Table 6**: Correct answers for 10 students who participated in all three activities (percentages only)

| | Activity 1 Percentage correct | | Activity 2 Percentage correct | | Activity 3 Percentage correct | |
|---|---|---|---|---|---|---|
| Student | *Direction* | *Polarity* | *Direction* | *Polarity* | *Direction* | *Polarity* |
| 1 | 0.00% | 46.88% | 50.00% | 12.50% | 87.50% | 62.50% |
| 2 | 9.38% | 9.38% | 37.50% | 87.50% | 100.00% | 100.00% |
| 3 | 0.00% | 40.63% | 75.00% | 100.00% | 100.00% | 62.50% |
| 4 | 0.00% | 43.75% | 75.00% | 87.50% | 100.00% | 100.00% |
| 5 | 53.13% | 68.75% | 87.50% | 75.00% | 100.00% | 87.50% |
| 6 | 40.63% | 59.38% | 62.50% | 100.00% | 62.50% | 87.50% |
| 7 | 12.50% | 43.75% | 87.50% | 0.00% | 62.50% | 0.00% |
| 8 | 18.75% | 34.38% | 87.50% | 87.50% | 100.00% | 100.00% |
| 9 | 50.00% | 53.13% | 62.50% | 100.00% | 100.00% | 100.00% |
| 10 | 31.25% | 18.75% | 100.00% | 75.00% | 50.00% | 0.00% |
| Mean | 21.56% | 41.88% | 72.50% | 72.50% | 86.25% | 70.00% |
| Median | 15.63% | 43.75% | 75.00% | 87.50% | 100.00% | 87.50% |
| St. Dev. | 20.80% | 17.75% | 19.36% | 36.23% | 19.94% | 39.62% |
| CV | 0.96 | 0.42 | 0.27 | 0.50 | 0.23 | 0.57 |

Figure 12b shows that there is persistent variability in how accurately these 10 students can identify polarity of causal links. In fact, the graphs for coefficients of variation in Figure 12c and the corresponding numbers in Table 6 suggest that the variation in their results for polarity increased from activity to activity. However, their results became more focused for direction classification as they progressed through the exercises.

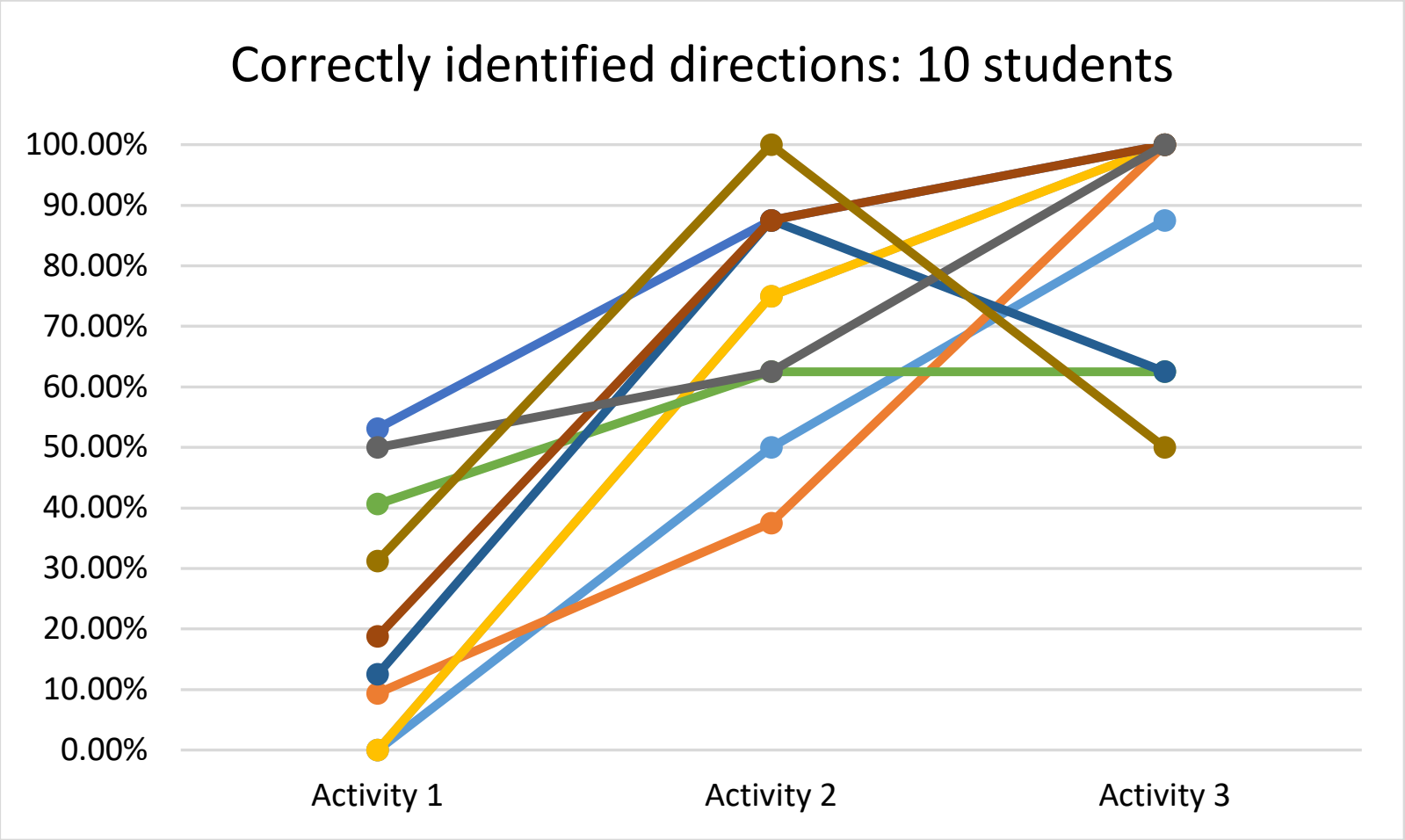

(a)

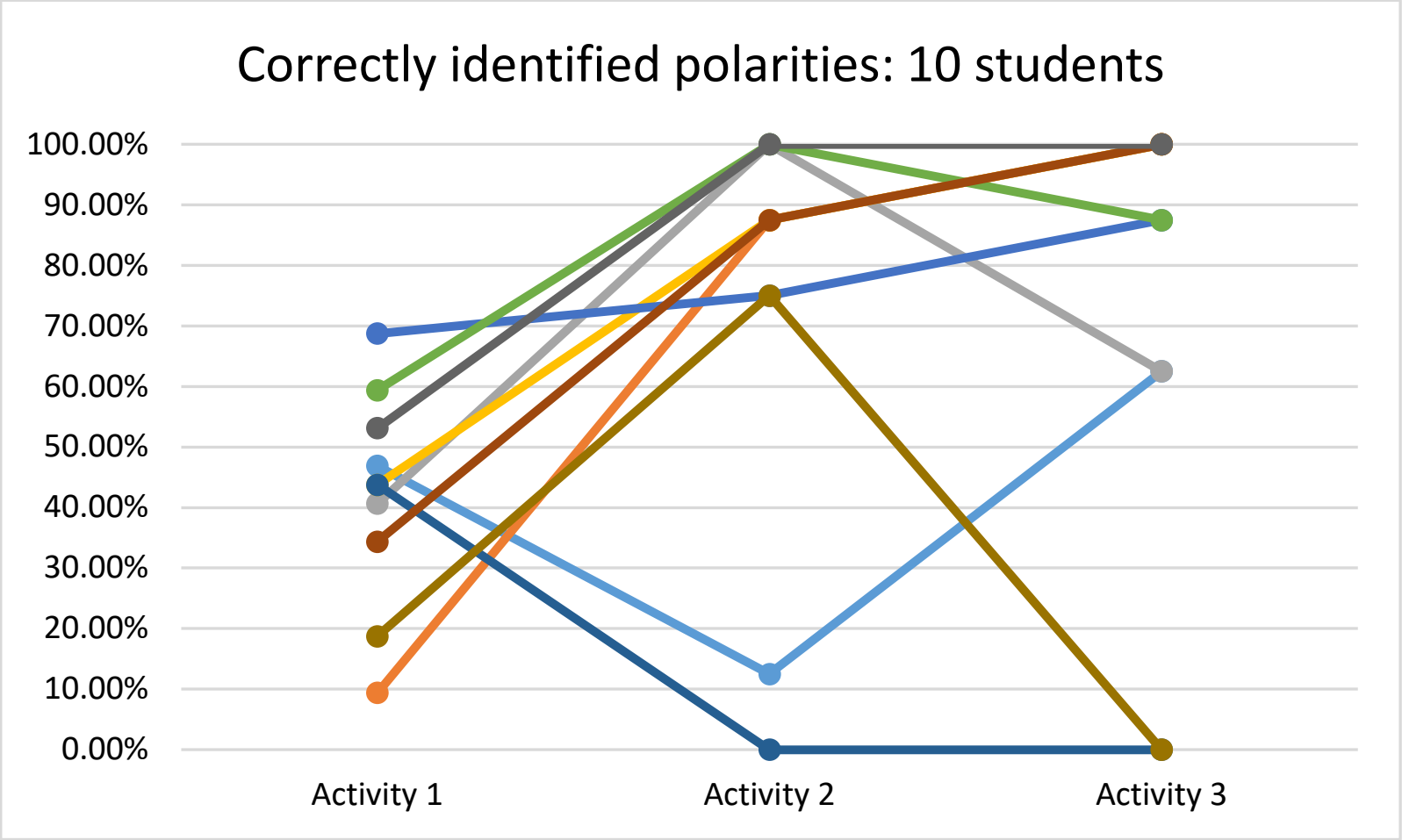

(b)

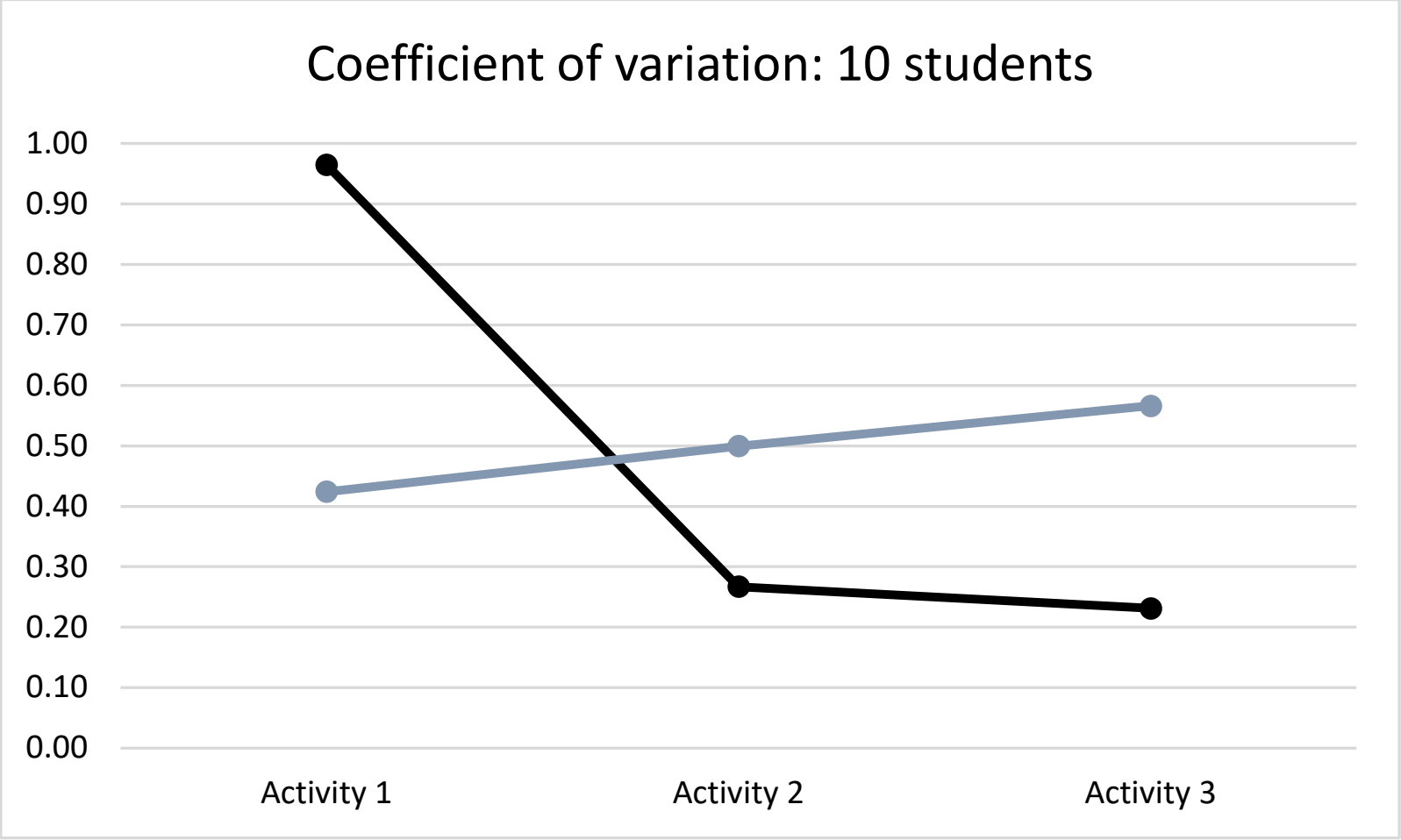

(c )

**Figure 12:** Statistics for 10 students

In summary, the improvements in students' ability to identify correctly causal directions and polarities from Activity 1 to Activity 2 suggests that using a smaller causal diagram and including relevant passages from the textbook in the worksheet were helpful strategies. However, it became clear that some students still struggled with constructing causal diagrams. Even after being shown the solution in Activity 2, several students were unable to accurately recreate the same diagram in Activity 3.

## Conclusion

This research examined the use of *causal diagrams* and *structural debriefing* to support *literacy-targeted instruction* in macroeconomics. Instructors who follow the literacy-targeted approach prioritize in-depth explorations of fewer concepts. Causal diagrams help students process complex economic information by constructing and modifying schemas that represent economic material. By integrating causal diagrams with conventional textbook explanations, instructors offer students varied perspectives that can enhance their understanding of economic theories. Structural debriefing is a pedagogical activity designed to introduce causal diagrams into the traditional classroom.

When used alongside conventional macroeconomics textbooks, structural debriefing proved effective, as evidenced by students' progress through successive activities. Our experiments revealed that while students generally improved in identifying cause-and-effect relationships in economic models, the concept of causal polarity remained challenging. Classroom observations and student feedback highlighted the need for clearer explanations of causal notation. Future studies should aim to refine the use of causal diagrams and structural debriefing as an instructional approach to better support students' comprehension of economic texts.

## Disclosure statement

The authors report there are no competing interests to declare.

## References

Ambrose, S. A., M. W. Bridges, M. DiPietro, M.C. Lovett and M.K. Norman (2010). How learning works: Seven research-based principles for smart teaching. Hoboken, NJ: John Wiley and Sons.

Capelo, C. and A. Silva (2020). "Optimising the Learning Potential of Simulations Through Structural Transparency and Exploratory Guidance." Simulation & Gaming: 104687812091620.

Capelo, C., R. Pereira and J. F. Dias (2021). "Teaching the dynamics of the growth of a business venture through transparent simulations." The International Journal of Management Education **19**(3): 100549. https://www.sciencedirect.com/science/article/pii/S1472811721000987.

Capelo, C., R. Pereira and J. F. Dias (2024). "Expanding model transparency and learning potential through structural and behavioural debriefings." Systems Research and Behavioral Science **n/a**(n/a). https://doi.org/10.1002/sres.3045.

Cavana RY, Dangerfield BC, Pavlov OV, Radzicki MJ and Wheat ID, Eds. 2021. Feedback Economics: Economic Modeling with System Dynamics. Springer: New York.

Collins, A., Joseph, D., & Bielaczyc, K. (2004). Design research: Theoretical and methodological issues. The Journal of the Learning Sciences, 13(1), 15–42.

Crookall, D. (2010). Serious games, debriefing, and simulation/gaming as a discipline. Simulation & Gaming: An International Journal, 41(6), 898 – 920.

Goffe, L and Wolla, S. (2024). Cognitive science teaching strategies and literacy-targeted economics complementarities. The Journal of Economic Education, 55-2, 156-165.

Grotzer, T. A. (2012). Learning Causality in a Complex World: Understandings of Consequence. Lanham, MD, R&L Education.

Jägerskog A-S. 2021a. The Affordance of Visual Tools. The Potential of Visual Representations of Pricing Facilitating an Epistemic Practice in Economics Teaching. Journal of Social Science Education 20 (1): 65-90.

Jägerskog A-S. 2021b. Using Visual Representations to Enhance Students' Understanding of Causal Relationships in Price. Scandinavian Journal of Educational Research 65 (6): 986-1003. https://doi.org/10.1080/00313831.2020.1788146.

Jägerskog A-S, Davies P and Lundholm C. 2019. Students' Understanding of Causation in Pricing: A Phenomenographic Analysis. Journal of Social Science Education 18 (3). https://www.jsse.org/index.php/jsse/article/view/1421.

Kim, Y. J., Pavlov, O. V. (2019). Game-based structural debriefing How can teachers design game-based curricula for systems thinking? Information and Learning Sciences. 120 (9/10), 567-588. https://doi.org/10.1108/ILS-05-2019-0039

Lederman, L. C. (1992). Debriefing: Toward a systematic assessment of theory and practice. Simulation & Gaming: An International Journal, 23(2), 145-160.

Maani KE and Cavana RY. 2007. Systems Thinking, System Dynamics: Managing Change and Complexity. Pearson Education New Zealand: North Shore, N.Z.

Mankiw, N. Gregory. 2016. Macroeconomics, 9th edition, Worth Publishers, Macmillan Learning, New York.

Mankiw, N.G., 2020. The Past and Future of Econ 101: The John R. Commons Award Lecture. The American Economist, 66(1), 9-17.

McKenney, S. and Reeves, T.C. (2012), Conducting Educational Design Research, Routledge, London.

Pavlov, O. V., Y. J. Kim and C. Whitlock (2019). Food Fight: Teaching Systems Thinking and Ecosystems. Learning, Education, & Games, Volume 3: 100 Games to Use in the Classroom and Beyond. K. Schrier, ETC Press: 150-155. https://ssrn.com/abstract=3511479.

Pavlov, O. V., K. Saeed, and L. W. Robinson (2015). Improving Instructional Simulation with Structural Debriefing. Simulation and Gaming, 46 (3-4), 383–403.

Pearl, J. and D. Mackenzie (2018). The Book of Why: The New Science of Cause and Effect. New York, Basic Books.

Perkins, D. N. and T. A. Grotzer (2005). "Dimensions of Causal Understanding: the Role of Complex Causal Models in Students' Understanding of Science." Studies in Science Education 41: 117-166.

Qudrat-Ullah, H. (2020). "Improving Human Performance in Dynamic Tasks with Debriefing-Based Interactive Learning Environments: An Empirical Investigation." International Journal of Information Technology & Decision Making **19**(4): 1065-1089.

Sterman JD. (2000). Business Dynamics. McGraw-Hill: Boston, MA.

Sterman, J. D. (2010). "Does formal system dynamics training improve people's understanding of accumulation?" System Dynamics Review 26(4): 316-334. https://doi.org/10.1002/sdr.447.

## Appendix A: Activity 1 – The national income model, Questions

These are multiple-choice questions for Activity 1. Students were asked to answer these questions twice, before and after the causal diagram of the national income model was introduced. Correct answers are presented in bold.

Date _______________ Team #_____

Team Members:

________________________ ________________________

________________________ ________________________

Q1: Assuming in the short-term that labor and capital employed in the economy remain constant, the wide adoption of AI, which is a technological innovation, is likely to have the following effect on consumption in the short term.

**a) Consumption will increase**
b) Consumption will decrease
c) AI will have no effect on consumption
d) This effect cannot be determined
e) I have no idea

Q2: An increase in government purchases is likely to have the following effect on the interest rate.

**a) Increase**
b) Decrease
c) Have no effect
d) Cannot be determined
e) I have no idea

Q3: A higher interest rate is likely to have the following effect on private savings.

**a) Increase**
b) Decrease
c) Have no effect
d) Cannot be determined
e) I have no idea

## Appendix B: Activity 2 - The government-purchases multiplier

This appendix includes the forms that were distributed to students during the structural debriefing of the multiplier effects.

**FORM 1**
**Government-purchases multiplier**

Date: ____________________ Instructor _________________
Team # ____________ Course: __________________

Team Members:

________________________ ________________________
________________________ ________________________

Please read the following quotes from Mankiw' Macroeconomics (9th ed) that explain the nature of the government-purchases multiplier (pp. 314-317):

> *Assuming that the economy is closed, so that net exports are zero, we write planned expenditure PE as the sum of consumption C, planned investment I, and government purchases G:*
>
> $$PE = C + I + G$$
>
> ***An Increase in Government Purchases in the Keynesian Cross****: An increase in government purchases of raises planned expenditure by that amount …. (Source: caption for Figure 11-5, p. 317).*
>
> *…according to the consumption function $C = C(Y - T)$, higher income causes higher consumption. When an increase in government purchases raises income, it also raises consumption, which further raises income, which further raises consumption, and so on. Therefore, in this model, an increase in government purchases causes a greater increase in income.*
>
> *How big is the multiplier? To answer this question, we trace through each step of the change in income. The process begins when expenditure rises by $\Delta G$, which implies that income rises by $\Delta G$ as well. This increase in income in turn raises consumption by $MPC \times \Delta G$, where MPC is the marginal propensity to consume. This increase in consumption raises expenditure and income once again. This second increase in income of $MPC \times G$ again raises consumption, this time by $MPC \times (MPC \times \Delta G)$, which again raises expenditure and income, and so on. This feedback from consumption to income to consumption continues indefinitely.*

**Instructions:** Using the provided text, complete the 'causal skeleton' graph that explains the mechanism of the government-purchases multiplier. Show causal directions and causal polarities of the edges. Positive causal relationships between variables are drawn as positive arrows. Negative links show inverse causal relationships. Please also identify the polarity of the feedback loop. Is it a positive (reinforcing) or negative (balancing) feedback loop?

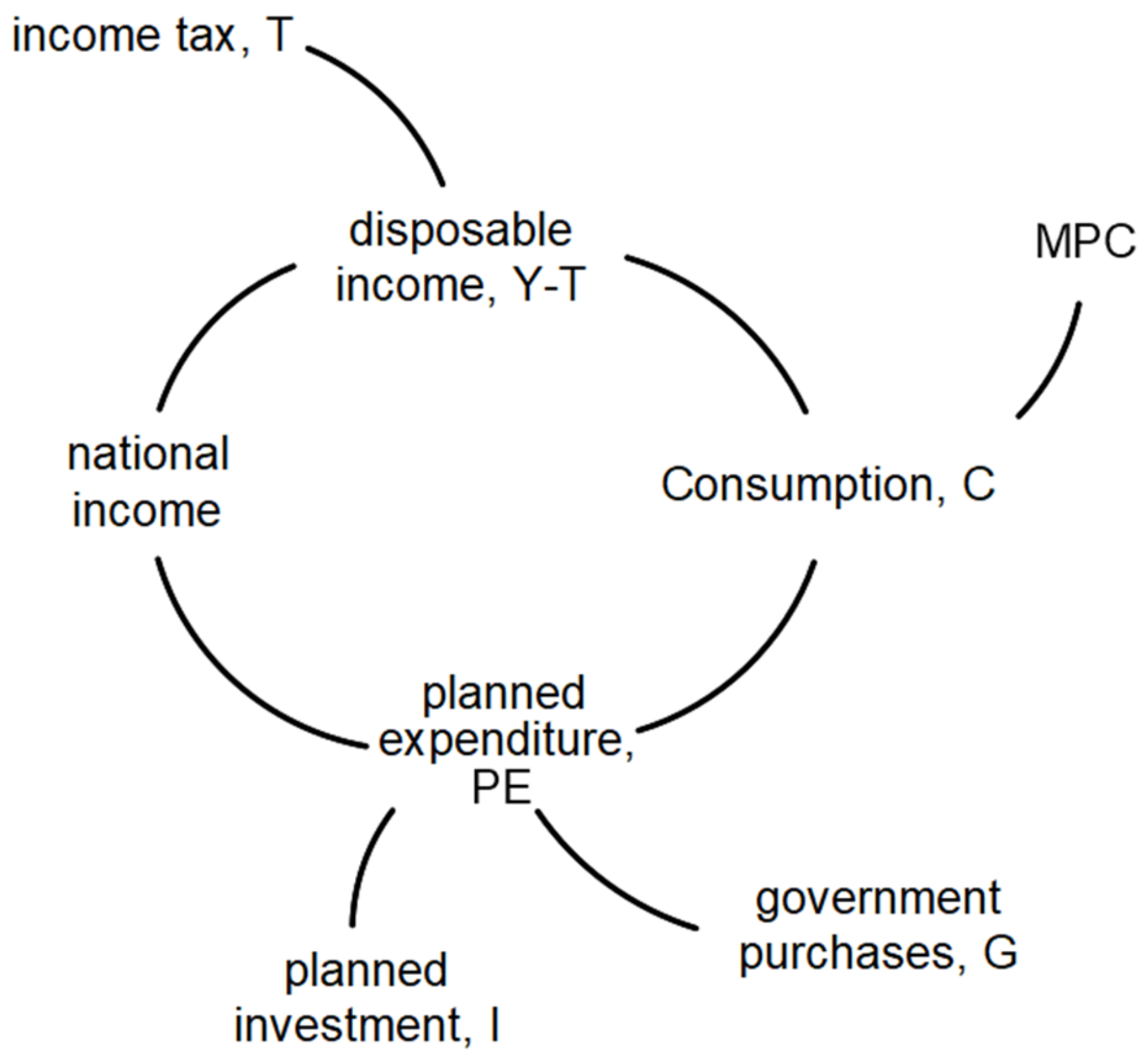

## Appendix C: Activity 3 – The tax multiplier

**FORM 2**
**Tax multiplier**

Date: ____________________ Instructor _________________
Team # ____________ Course: ____________________

Team Members:
_______________________ _______________________
_______________________ _______________________

Please read the following quote from Mankiw's Macroeconomics (9th ed) that explains the nature of the tax multiplier (pp. 318-319):

> *A decrease in taxes of $\Delta T$ immediately raises disposable income $Y - T$ by $\Delta T$ and, therefore, increases consumption by $MPC \times \Delta T$. For any given level of income Y, planned expenditure is now higher… Just as an increase in government purchases has a multiplied effect on income, so does a decrease in taxes. As before, the initial change in expenditure, now $MPC \times \Delta T$, is multiplied by $1/(1 - MPC)$.*

**Instructions:**

1. Using the above text, complete the graph below that explains the mechanism of the tax multiplier. Show causal directions and causal polarities of the edges. Positive causal relationships between variables are drawn as positive arrows. Negative links show inverse causal relationships. Please also identify the polarity of the feedback loop. Is it a positive (reinforcing) or negative (balancing) feedback loop?
2. Use the causal diagram to explain the tax multiplier effect of a decrease in taxes, ΔT.

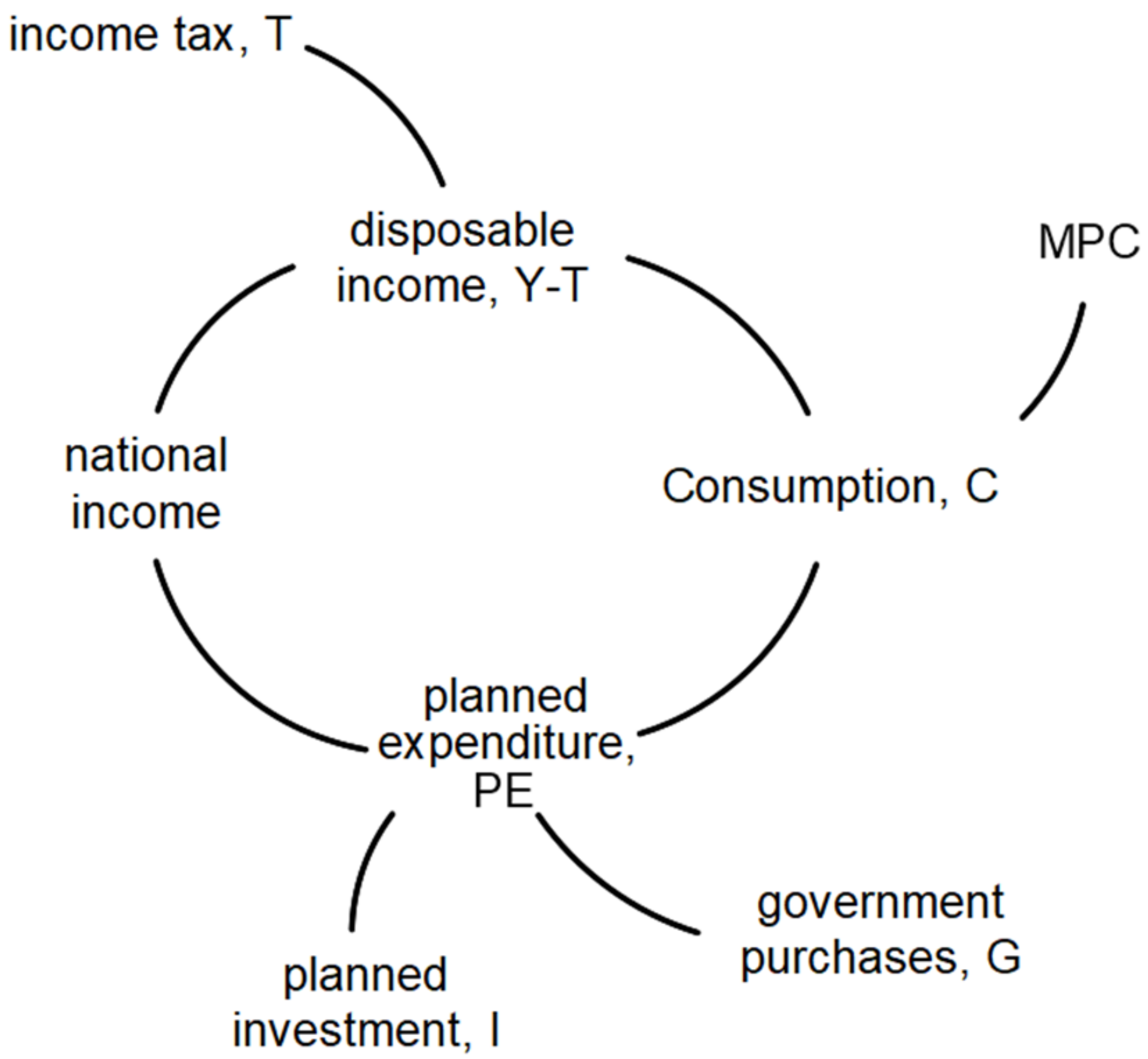

## Appendix D: Form 3 – Questionnaire

**FORM 3**
**G & T multipliers**

Date: ____________________ Instructor _________________
Name: ____________________ Course: ____________________

Now that you've reviewed the government purchases multiplier and the tax multiplier, please answer the following questions.

1. Are there any **similarities** between the economic structures that are responsible for the government purchases multiplier and the tax multiplier?

2. Are there any **differences** between the economic structures that are responsible for the government purchases multiplier and the tax multiplier?

3. Were the causal diagrams helpful in understanding the multiplier effects? Please elaborate.

4. Did you find any part of the activity with causal diagrams confusing?

5. Do you think you can apply the causal diagrams approach to explain situations that you face at work, at home, read about in the news, etc.? Can you provide an example?

## Appendix E: Instructor Manual

# Instructor Manual

Version: 2.0
October 22, 2024

Oleg V. Pavlov
Natalia V. Smirnova
Elena V. Smirnova

This is an activity called "structural debriefing" [4] that visualizes the logical structure of the models presented in Mankiw's Macroeconomics. In this analysis, we're not arguing whether or not the models presented in Mankiw's Macroeconomics are accurate.

Note: all the pages refer to Mankiw's Macroeconomics, 9th edition.

In this section we conduct structural debriefing of the Mankiw explanations for government-purchase multiplier and tax multiplier.

## Government Purchases-Multiplier

Mankiw explains the multiplier effects. First, Mankiw defines planned expenditure (p. 314):

> *Assuming that the economy is closed, so that net exports are zero, we write planned expenditure PE as the sum of consumption C, planned investment I, and government purchases G:*
>
> $$PE = C + I + G$$

Then Mankiw makes the following observation:

> ***An Increase in Government Purchases in the Keynesian Cross****: An increase in government purchases of $\Delta G$ raises planned expenditure by that amount .... (Source: caption for Figure 11-5, p. 317).*

Here is a quote from Mankiw (p. 317) that explains the government-purchases multiplier model:

> *...according to the consumption function $C = C(Y - T)$, higher income causes higher consumption. When an increase in government purchases raises income, it also raises consumption, which further raises income, which further raises*

[4] Pavlov, O. V., K. Saeed and L. W. Robinson (2015). "Improving Instructional Simulation with Structural Debriefing." Simulation & Gaming 46(3-4): 383-403.

*consumption, and so on. Therefore, in this model, an increase in government purchases causes a greater increase in income.*

*How big is the multiplier? To answer this question, we trace through each step of the change in income. The process begins when expenditure rises by* $\Delta G$*, which implies that income rises by* $\Delta G$ *as well. This increase in income in turn raises consumption by* $MPC \times \Delta G$*, where MPC is the marginal propensity to consume. This increase in consumption raises expenditure and income once again. This second increase in income of* $MPC \times \Delta G$ *again raises consumption, this time by* $MPC \times (MPC \times \Delta G)$*, which again raises expenditure and income, and so on. This feedback from consumption to income to consumption continues indefinitely.*

## *Task*

Form student teams. Ask each team to complete the graph in Figure 1, which is called a "causal skeleton"[5], using the description above from Mankiw. Show causal directions between variables by adding arrow heads to the edges. Indicate whether the causal relationships are positive (show a plus, +) or negative (show a minus, -). Please also identify the feedback loop and its polarity. Is it a positive (reinforcing) or negative (balancing) feedback loop?

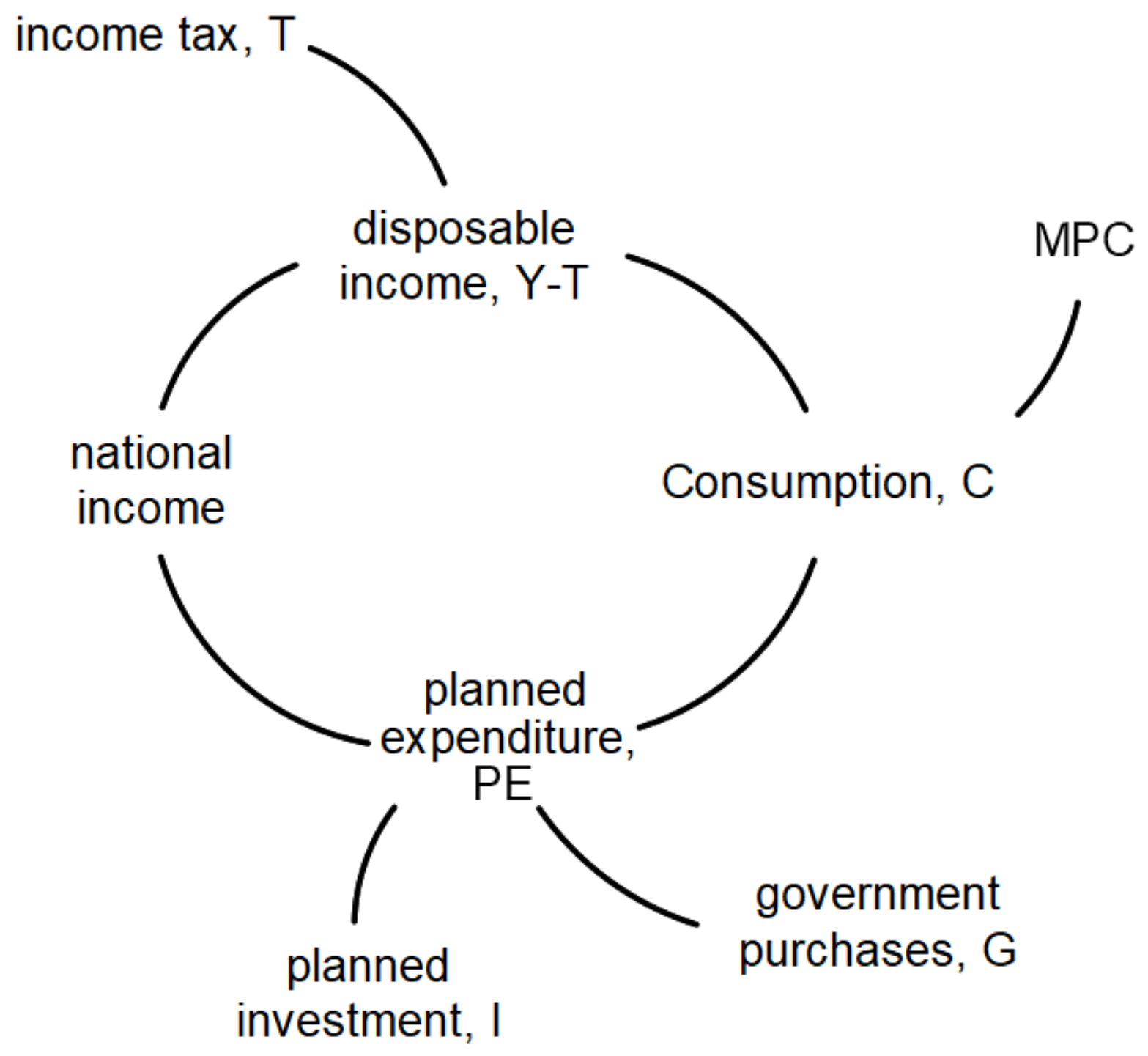

[5] The causal skeleton is an undirected graph that shows causal relationships between variables.

**Figure 1**: Causal skeleton that shows the logic of the government-purchases multiplier.

**Solution**

The solution to the above question is shown in Figure 2. There is only one inverse relationship between taxes and disposable income. All other causal relationships are positive. Since there are only positive arrows in the loop, the feedback loop is reinforcing, which is shown by placing a plus sign in the center of the loop. A curved arrow shows the direction of the feedback loop.

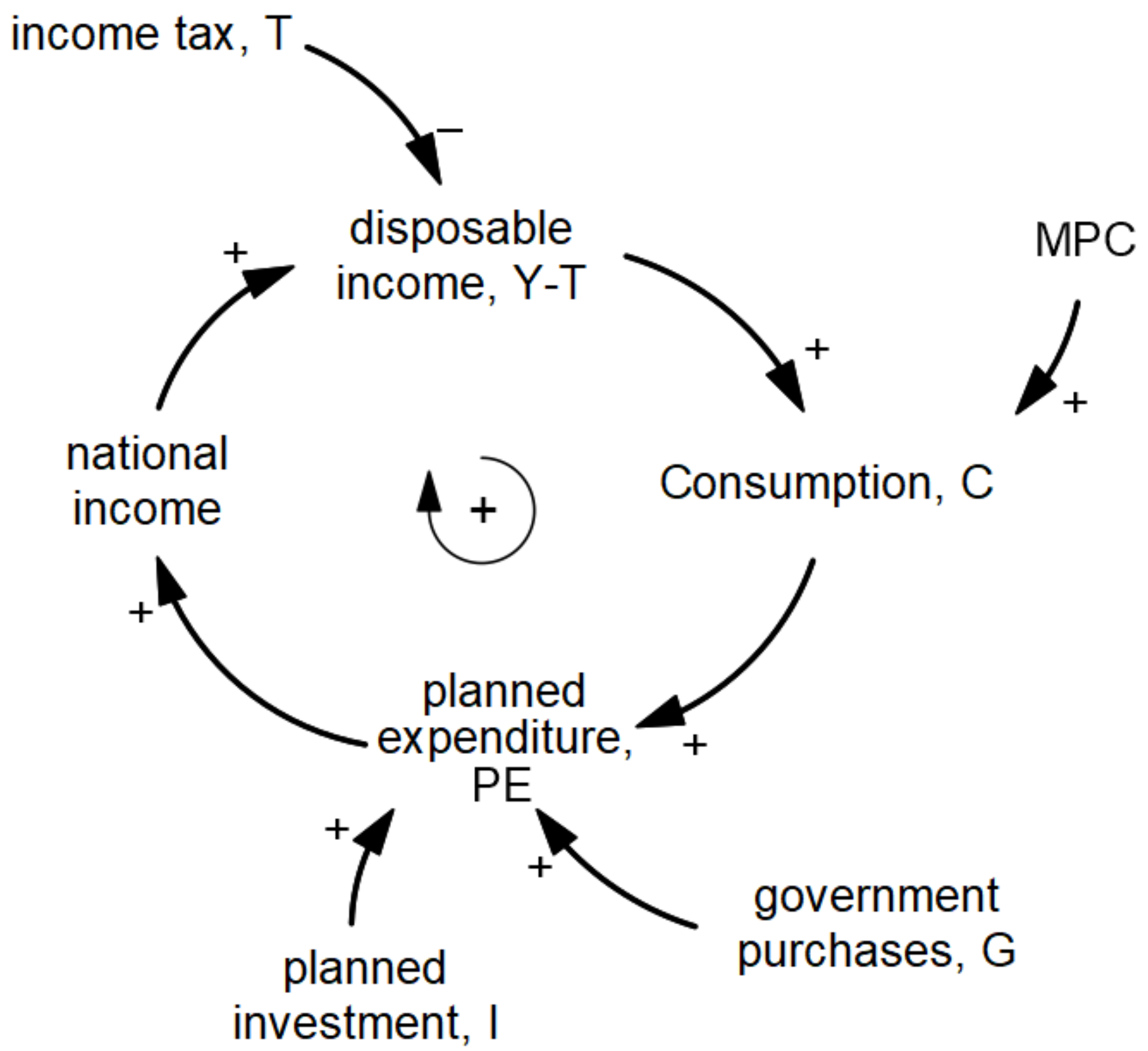

**Figure 2**: Causal diagram that shows the logic of the government-purchases multiplier. This figure corresponds to the causal skeleton in Figure 1.

Figure 3 explains how the initial increase in government purchase ΔG can be traced around the loop as it adds to the consumption smaller amounts with each iteration. Start the analysis by imaging that government purchases increase by ΔG. The size of each red arrow indicates the relative increase of the variable value during an iteration around the feedback loop. Note that during the first iteration,

C increases by MPC x ΔG, which is less that ΔG; therefore, the first consumption arrow is shorter than the arrow for ΔG next to the variable government purchases.

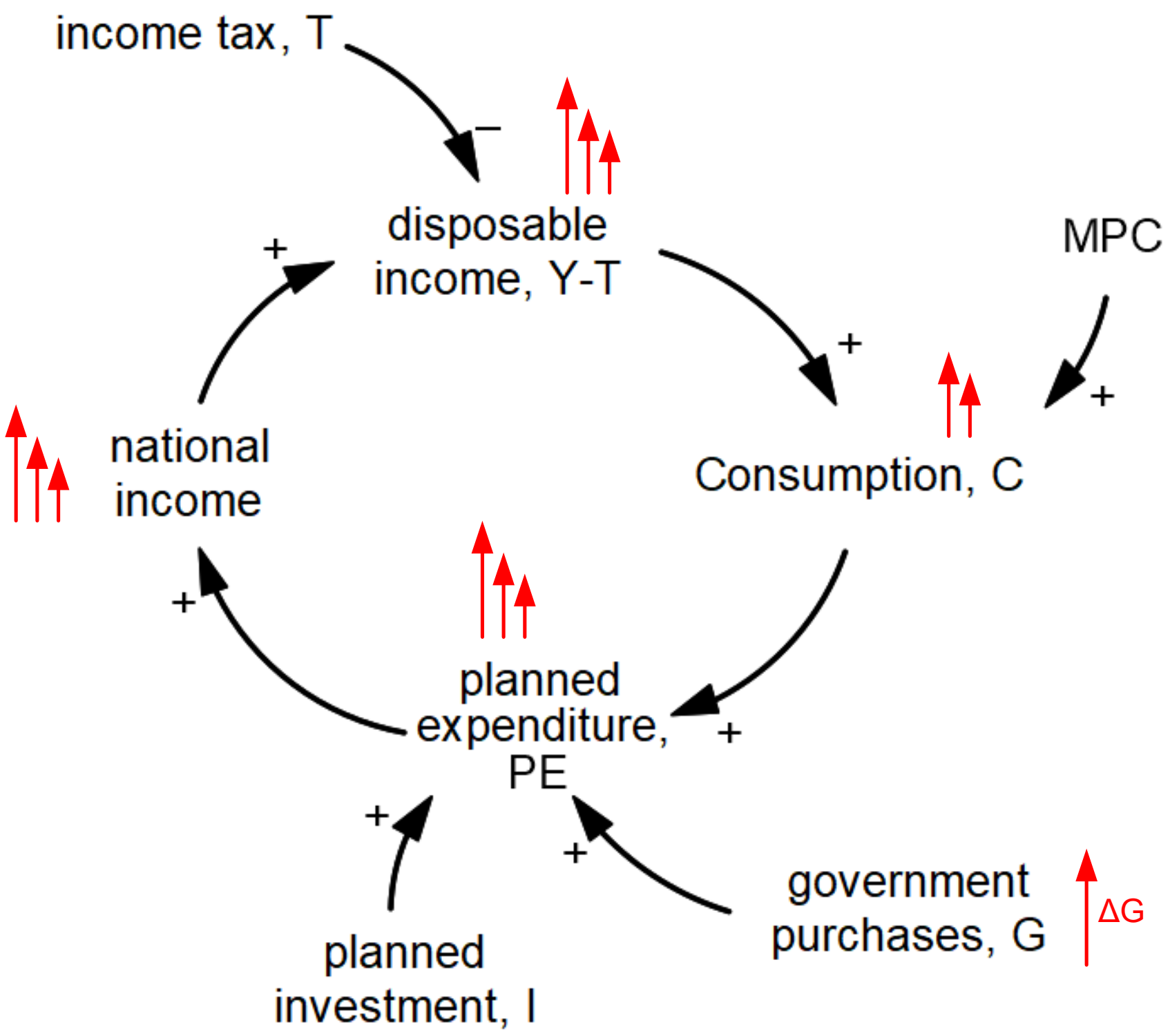

**Figure 3**: Government purchases-multiplier effect.

Confirm Mankiw's (p. 317) explanation of the total effect on **national income**, ΔY, that is

| | |
|---|---|
| Initial Change in Government Purchases | $= \Delta G$ |
| First Change in Consumption | $= MPC \times \Delta G$ |
| Second Change in Consumption | $= MPC^2 \times \Delta G$ |
| Third Change in Consumption | $= MPC^3 \times \Delta G$ |
| . | . |
| . | . |
| . | . |

$$\Delta Y = (1 + MPC + MPC^2 + MPC^3 + \ldots)\, \Delta G$$

Then, the **government-purchases multiplier** is:

$$\Delta Y / \Delta G = (1 + MPC + MPC^2 + MPC^3 + \ldots)\, \Delta G / \Delta G$$

$$= 1 + MPC + MPC^2 + MPC^3 + \ldots$$ *an infinite geometric series*

$= 1 / (1 - MPC)$

## Tax Multiplier

The above graphs can also be used to explain the logic of the tax multiplier. Here is a quote from Mankiw's Macroeconomics (9th ed) that explains the nature of the tax multiplier (pp. 318-319):

> *A decrease in taxes of $\Delta T$ immediately raises disposable income $Y - T$ by $\Delta T$ and, therefore, increases consumption by $MPC \times \Delta T$. For any given level of income Y, planned expenditure is now higher… Just as an increase in government purchases has a multiplied effect on income, so does a decrease in taxes. As before, the initial change in expenditure, now $MPC \times \Delta T$, is multiplied by $1/(1 - MPC)$.*

**Task**

Ask the teams that have been formed earlier to fill out the causal skeleton in Figure 4. It is identical to the causal skeleton in Figure 2 since the tax multiplier works the same way as a government purchase injection.

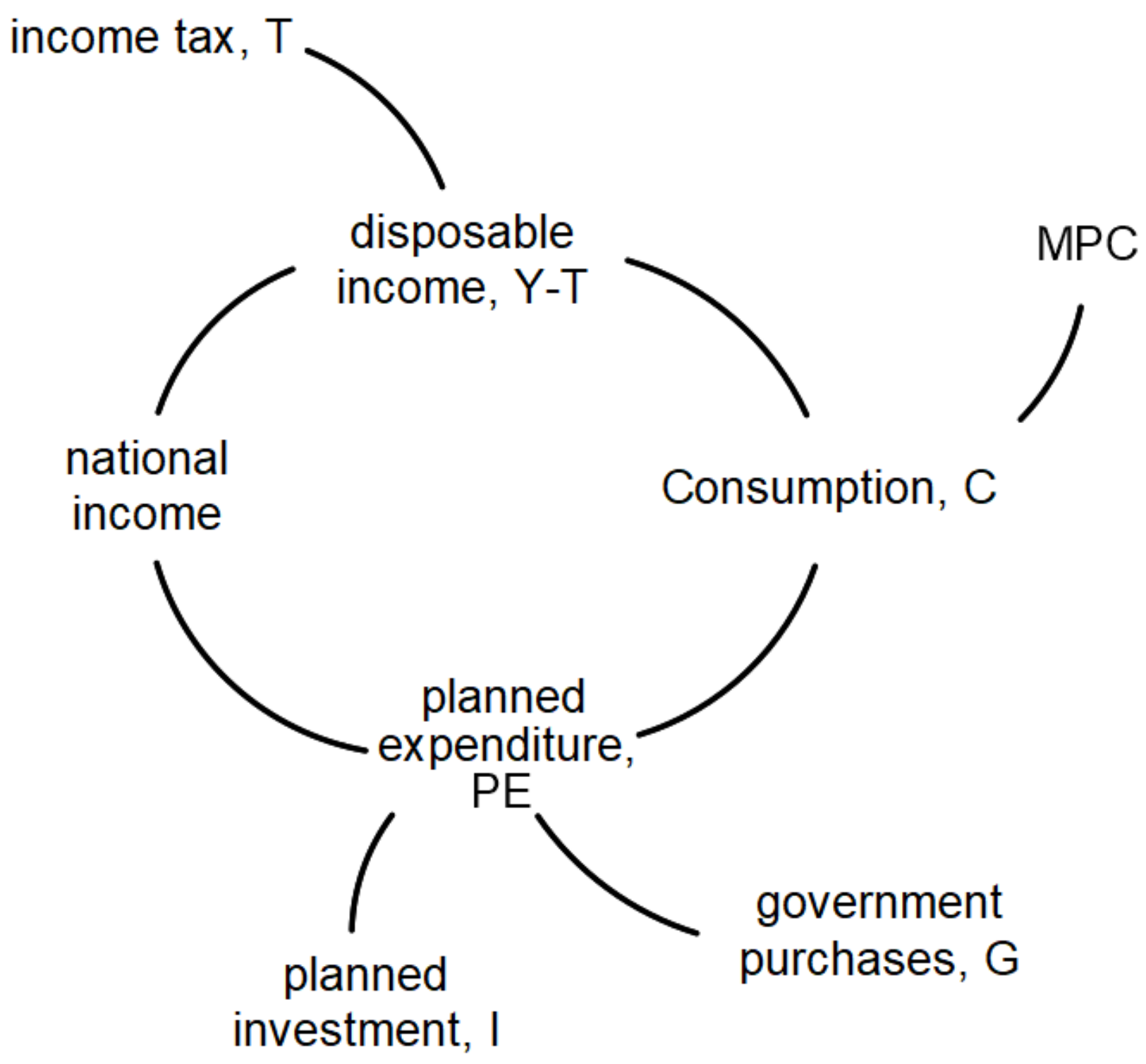

**Figure 4**: Causal skeleton that shows the logic of the government-purchase multiplier.

**Solution**

The tax multiplier mechanism is shown in Figure 5. It is identical to the government-purchase multiplier mechanism in Figure 2.

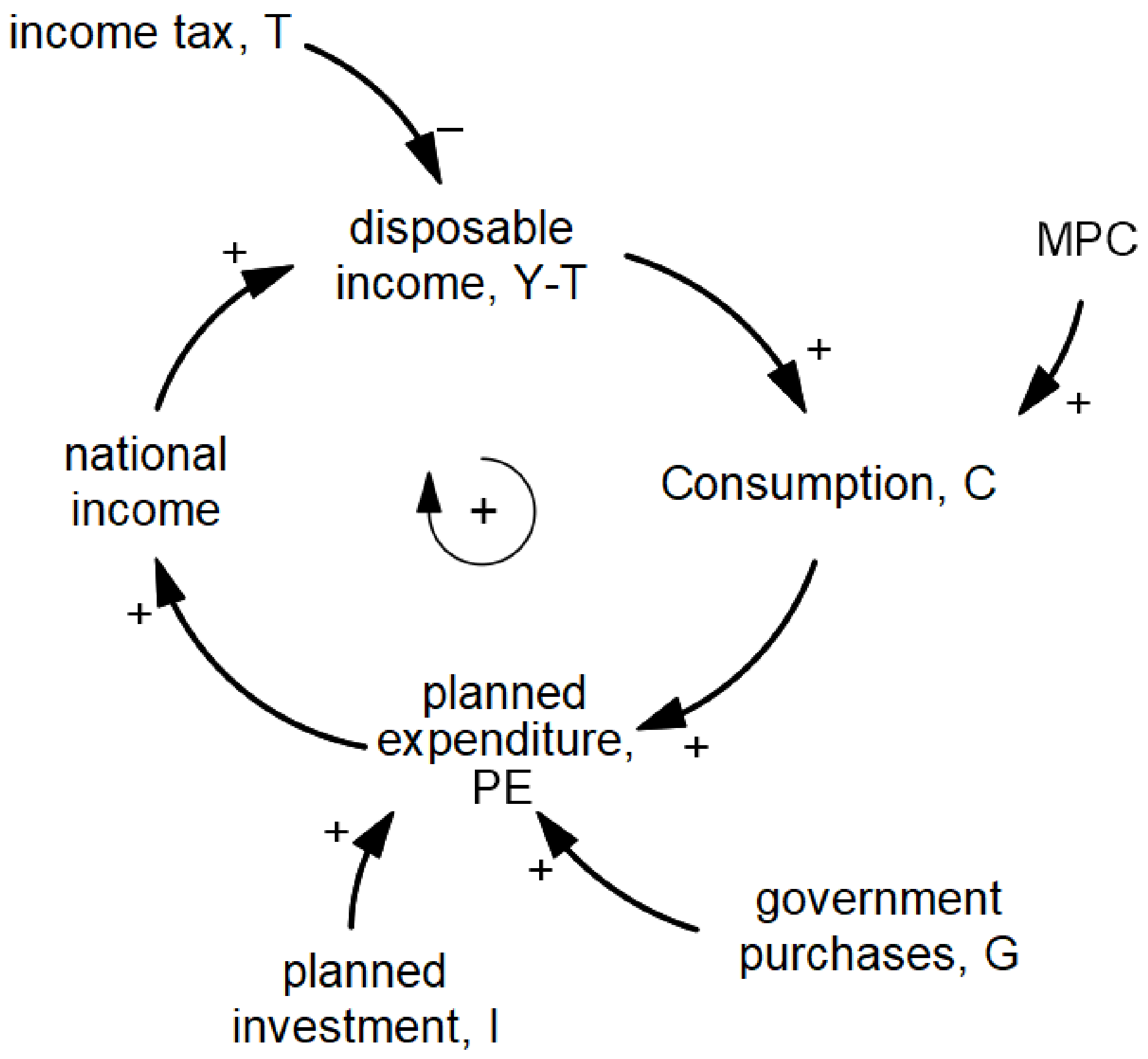

**Figure 5**: Causal diagram that shows the logic of the tax multiplier.

Figure 6 explains the tax multiplier effect of a tax decrease. A decrease in taxes, ΔT, increases the disposable income by ΔT, which increases consumption and planned expenditure by $MPC \times \Delta T$. The second increase in consumption and planned expenditure is $MPC \times (MPC \times \Delta T)$.

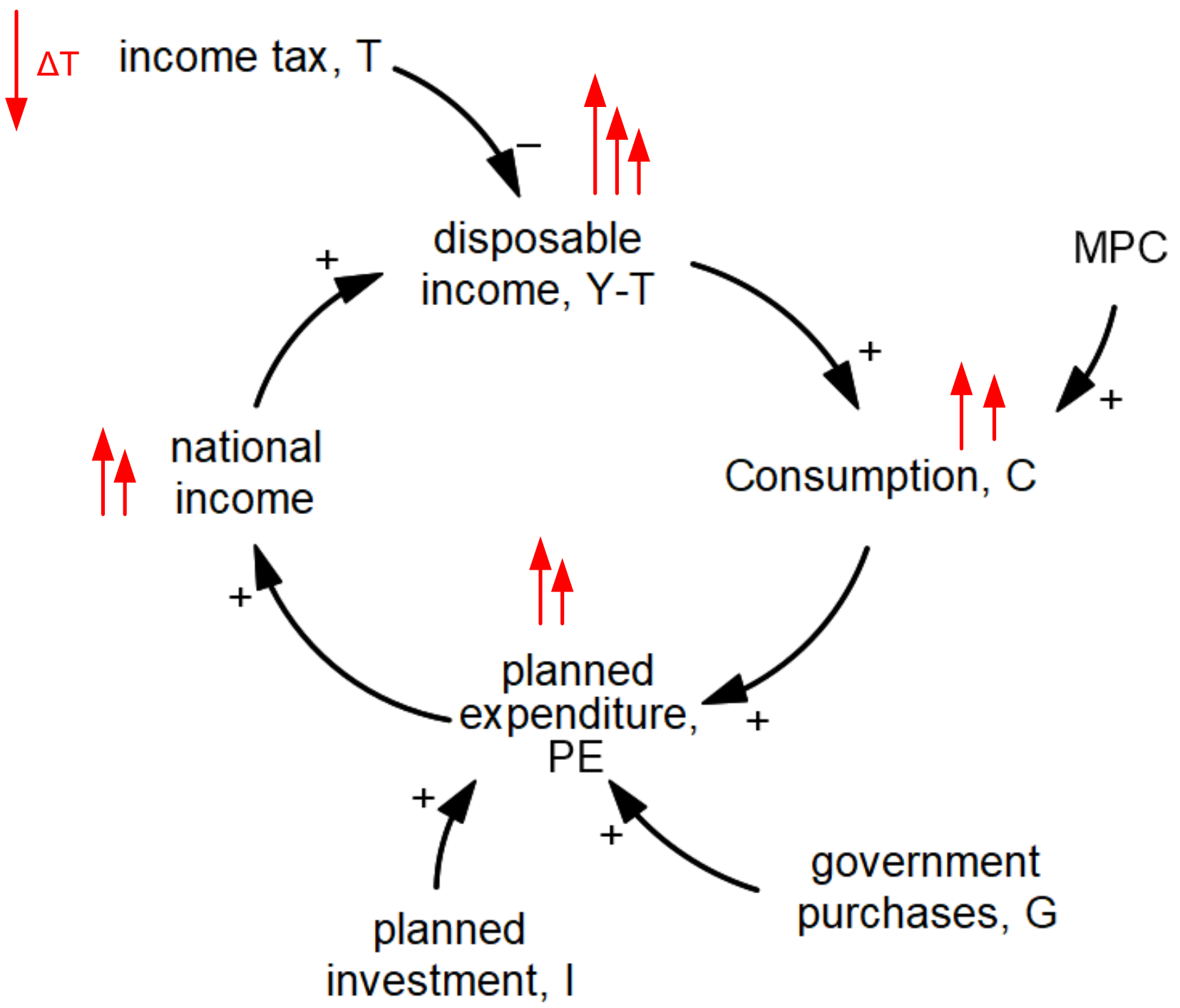

**Figure 6**: The tax-multiplier effect following a decrease in taxes, ΔT.

The total effect on **national income**, ΔY, is

| | |
|---|---|
| Change in Taxes | $= -\Delta T$ |
| Change in Disposable Income | $= \Delta T$ |
| First Change in Consumption | $= MPC \times \Delta T$ |
| Second Change in Consumption | $= MPC^2 \times \Delta T$ |
| Third Change in Consumption | $= MPC^3 \times \Delta T$ |
| . | . |
| . | . |
| . | . |

$$\Delta Y = (MPC + MPC^2 + MPC^3 + \dots)\,\Delta T$$

The **tax multiplier** is:

$$\Delta Y / \Delta T = (MPC + MPC^2 + MPC^3 + \dots)\,\Delta T / \Delta T$$

$$= MPC + MPC^2 + MPC^3 + \dots$$ *an infinite geometric series*

$= MPC ( 1 + MPC + MPC^2 + MPC^3 + \ldots )$

$= MPC ( 1 / (1 - MPC ) )$

$= MPC / (1 - MPC )$

## Possible modifications

**Option 1**

The two tasks above can be performed independently as two activities.

**Option 2**

Because the two tasks are very similar, a possible way to cover the material about the multipliers is to split the class into teams, and then ask some teams to analyze the government-purchases multiplier and the other teams can analyze the tax multiplier. Then the instructor can ask the teams to compare notes and discuss how the mechanisms of those multipliers are similar and how they are different. Below we show the similarities and differences:

Similarities in the analysis:

- The same economic structure is responsible for both multipliers: directions of causal relationships and polarities between variables are the same in both cases (Figure 2 and Figure 5 are identical).
- In both cases, there is a positive feedback loop (the same reinforcing loop in Figure 2 and Figure 5).

Differences in the analysis:

- In Figure 3, government purchases increase by $\Delta G$ and then planned expenditure increase by the same amount $\Delta G$.
- In Figure 6, taxes decrease by $\Delta T$, but then disposable income increases by same amount $\Delta T$.